\documentclass[11pt,a4paper]{article}
\usepackage{float}
\usepackage{diagbox}
\usepackage{jheppub}
\usepackage{graphicx}
\usepackage{amsmath}
\usepackage{amssymb}
\usepackage{amsthm}
\usepackage{mathtools}
\usepackage{mathrsfs}
\input epsf
\usepackage{epsfig}
\usepackage{tikz}
\usepackage{subfigure,tikz}
\usetikzlibrary{backgrounds,automata}
\usepackage{graphics}
\usepackage{subfigure}
\usepackage[active]{srcltx}
\usepackage[T1]{fontenc} 
\usepackage{graphicx}

\newcommand{\bea}{\begin{eqnarray}}
	\newcommand{\eea}{\end{eqnarray}}
\newcommand{\bean}{\begin{eqnarray*}}
	\newcommand{\eean}{\end{eqnarray*}}
\newcommand{\nn}{\nonumber \\}

\def\W #1{\widetilde{#1}}

\def\abs#1{\left| #1\right|}

\def\co{\,,}
\def\ed{\,.}

\def\det{\mathop{\rm det}}

\def\eref#1{(\ref{#1})}

\def\d{{\rm d}}

\def\a{{\alpha}}

\def\b{{\beta}}

\def\d{\partial}

\def\eps{\epsilon}

\def\what{\widehat}
\newcommand{\ctoself}[2]{C^{({#1})}_{{#2}\to {#2}}}
\newcommand{\ctonext}[3]{C^{({#1})}_{{#2}\to {#2};\widehat{{#3}}}}

\def\Label#1{\label{#1}%
	\smash{\hbox to0pt{\raise1ex\hbox{\tiny[#1]}\hss}}}

\newcommand{\parall}[2]{{#1}\ /\kern -0.8em / \  {#2}}

\title{\boldmath  Reduction with Degenerate Gram matrix  for One-loop Integrals}

\author[a,b,d,e]{Bo Feng,}
\author[c,f]{Chang Hu,}
\author[a]{Tingfei Li,}
\author[a]{Yuekai Song}

\affiliation[a]{Zhejiang Institute of Modern Physics, Zhejiang University, Hangzhou, 310027, P. R. China }
\affiliation[b]{ Beijing Computational Science Research Center, Beijing 100084, China}
\affiliation[c]{Hangzhou Institute of Advanced Study, UCAS, Hangzhou, 310027, P.R. China}
\affiliation[d]{Center of Mathematical Science, Zhejiang University, Hangzhou, 310027, P. R. China}
\affiliation[e]{Peng Huanwu Center for Fundamental Theory, Hefei, Anhui, 230026, China}
\affiliation[f]{University of Chinese Academy of Sciences, 100190 Beijing, China}

\emailAdd{fengbo@zju.edu.cn}
\emailAdd{isiahalbert@126.com}
\emailAdd{tfli@zju.edu.cn}
\emailAdd{songyk@zju.edu.cn}

\abstract{An improved PV-reduction (Passarino-Veltman) method for 
	one-loop integrals with auxiliary vector $R$ has been proposed {in} \cite{Feng:2021enk,Hu:2021nia}. It has also been shown that the new method is a {self-completed} method in \cite{Feng:2022uqp}. Analytic reduction coefficients can be easily produced {by recursion relations in} this method, where the Gram determinant appears in denominators. The singularity caused by Gram determinant is a well-known fact and it is important to {address these divergences} in a given frame. In this paper, we propose a systematical algorithm to deal with this problem in our method. The key idea is that now the master integral of the highest topology will be decomposed into combinations of master integrals of lower topologies. By demanding the cancellation of divergence for obtained {general reduction coefficients},  we solve decomposition coefficients as a Taylor series of the Gram determinant. Moreover, the same idea can be applied to other kinds of divergences.

}


\begin{document}
	\maketitle
	\flushbottom

	\section{Introduction}
	\label{sec:intro}
	
	The calculation of scattering amplitudes at loop level is an important issue in High Energy Physics. A widely used method is to reduce a loop amplitude into a linear combination of some  scalar integrals under dimensional regularization~\cite{ Brown:1952eu,Melrose:1965kb,Passarino:1978jh,tHooft:1978jhc,vanNeerven:1983vr, Stuart:1987tt,vanOldenborgh:1989wn,Bern:1992em,Bern:1993kr,Fleischer:1999hq,Binoth:1999sp,Denner:2002ii,Duplancic:2003tv,Denner:2005nn,Ellis:2007qk,Ossola_2007, Bern:1994cg}
	\begin{align}
		I^{1-loop}=\sum_{i_{d_0+1}} C_{i_{d_0+1}} I_{d_0+1}^{i_{d_0+1}}+\sum_{i_{d_0}} C_{i_{d_0}} I_{d_0}^{i_{d_0}}+\cdots + \sum_{i_1} C_{i_1} I_1^{i_1}\co \label{reduction_d0}
	\end{align}
	where $i_s$ is the set of propagators appearing in the basis. The coefficient $C_{i_s}$ ($s=1,\cdots,d_0+1$) is simply a rational function of some Lorentz invariant contractions of external momenta as well as the masses and spacetime dimension, while the terms $I_s^{i_s}$ are the $s$-gon master integrals. With the general expansion \eref{reduction_d0}, the computation of general one-loop amplitudes  has been switched to determining those coefficients  $C_{i_s}$. Many tools have been invented to shovel the brambles, such as integration-by-parts (IBP) \cite{Chetyrkin:1981qh,Tkachov:1981wb}, PV (Passarino-Veltman) reduction~\cite{Passarino:1978jh}, OPP (Ossola-Papadopoulo-Pittau) reduction~\cite{Ossola:2006us,Ossola:2007bb,Ellis:2007br,Ossola_2007}, unitarity cut
	method, etc \cite{Bern:1994zx,Bern:1994cg,Britto:2004nc,Britto:2005ha,Campbell:1996zw,Bern:1997sc,Denner:2005nn,Anastasiou:2006gt,Britto:2010um}.
	
	All these methods can be divided into two categories, i.e., the reduction at the integrand level or the integral level. For reduction at the integrand level,
	\cite{Ossola_2007} shows how to extract the coefficients of the 4-, 3-, 2- and
	1-point one-loop
	scalar integrals from the full one-loop integrand of arbitrary scattering processes in an algebraical way. For the reduction at the integral level, an efficient way is the unitarity cut method. The main idea is to compare the imaginary part of two sides of \eref{reduction_d0}. However, since the loop-integral is well-defined only after using the dimensional regularization, the unitarity cut method in pure  4D needs to be generalized  to $(4-2\eps)$-dimension, which has been done in \cite{Anastasiou:2006jv,Anastasiou:2006gt}. Based on this generalization, the analytic expressions for reduction coefficients (except the tadpole coefficients) have been derived in a series of  papers \cite{Britto:2006fc,Britto:2007tt,Britto:2008vq,Britto:2008sw,Feng:2013sa}.

	Recently, we proposed a new framework for general one-loop tensor reduction by employing an auxiliary vector $R$ and two kinds of differential operators \cite{Feng:2021enk,Hu:2021nia}. Similar to other reduction methods, our method also suffers from divergences {for} vanishing Gram determinant, which appears as the inverse of Gram matrix in the recursion constructions of reduction coefficients.
	It is well known that the vanishing Gram determinant indicates external momenta are not completely independent of each other,  thus
	the master integral of the highest topology will not be 
	in the basis anymore, and it can be decomposed into combinations of master integrals of lower
	topologies.\footnote{In this paper, the integrals of \textbf{the highest topology} refer to the integrals containing the maximum number of  propagators while \textbf{lower topology} means the integrals containing fewer propagators, which is distinct from its definition in mathematics.}  With this decomposition, the reduction coefficients will be {reorganized so that} the divergences between different terms will cancel {with} each other. To see this picture 
	clearly, calculating analytic coefficients will be important, which is exactly the merit of our new method.  
	In this paper, we will systematically study the 
	degenerate case with vanishing Gram determinant using the results in \cite{Feng:2021enk,Hu:2021nia}. It turns out that the decomposition coefficients can easily be solved as
	a series in the Gram determinant. 
	
	This paper is organized as follows. In section \ref{sec:background}, we will review the improved PV-reduction method with auxiliary vector $R$  in our previous work {and  briefly discuss}  the general {structure} of reduction coefficients. In section \ref{sec:reduction}, based on the explicit computation for bubbles {in the limit} $K^2\to 0$, we establish an algorithm to deal
	with the vanishing Gram determinant for general topologies. We demonstrate the algorithm {with} the triangle in the main text and the box and pentagon in Appendix \ref{appdix:A}. In section \ref{sec:consistence}, an important
	consistency has been proved, i.e., our algorithm should be 
	independent of the auxiliary vector $R$ in reduction coefficients and the choice of tensor ranks. Finally, we give a summary and some discussions in section \ref{sec:discussion}.

	\section{Background}
	\label{sec:background}
	\subsection{Review of Improved PV-Reduction}

	In this section, we briefly review how to reduce  one-loop tensor integrals using our {\sl improved 
	PV-reduction method}. After that, we will explain how to deal with  the degenerated Gram determinant  in our framework.

	For a general one-loop $m$-rank tensor integral with $(n+1)$ propagators
	\bea
	I_{n+1}^{\mu_1\mu_2\cdots\mu_m}= \int {d^D \ell\over i\pi^{D/2}}  {\ell^{\mu_1}\ell^{\mu_2}... \ell^{\mu_m}\over D_0\prod_{j=1}^n D_j}=\int {d^D \ell\over i\pi^{D/2}}  {\ell^{\mu_1}\ell^{\mu_2}... \ell^{\mu_m}\over (\ell^2-M_0^2)\prod_{j=1}^n ((\ell-K_j)^2-M_j^2)}~~~~\label{set-1-1}
	\eea
	the algorithm of traditional PV (Passarino-Veltman) reduction method \cite{Passarino:1978jh} is following. First, 
	we use the external momenta to write down the matched general tensor structure. For example, with rank $3$ tensor
	integral of the form \eref{set-1-1}, we write down 
	\bea I_{n+1}^{\mu_1\mu_2\mu_3}=a_i(g^{\mu_1\mu_2}K_i^{\mu_3}+g^{\mu_1\mu_3}K_i^{\mu_2}+g^{\mu_2\mu_3}K_i^{\mu_1})+b_{ijl}K_i^{\mu_1}K_j^{\mu_2}K_l^{\mu_3} ~~~\label{new-I-1}\eea
	with unknown expansion coefficients $a_i, b_{ijk}$. Secondly, we contract both sides of \eref{new-I-1} with various $g^{\mu\nu}$ and $K_i$ to establish enough algebraic relations to solve these coefficients. It is easy to see that with higher and higher tensor rank, there will be more and more different tensor structures to be written down in \eref{new-I-1} and more and more algebraic relations to be established to fix them. 
	
	A nice observation made in \cite{Feng:2021enk,Hu:2021nia} is that the complicated tensor structure can be simply recovered from 
	\bea I^{(m)}_{n+1}\equiv\int {d^D \ell\over i\pi^{D/2}}  {(2R\cdot \ell)^m\over D_0\prod_{j=1}^n D_j}~~~~\label{set-1-2}
	\eea
	by writing $R=\sum_{i=1}^m x_i R_i$ and extracting terms with coefficients $x_1...x_m$ (or taking the proper differentiation action). For example, 
	\bea
	(\ell\cdot P)\ell^2\sim (\d_R\cdot \d_R)(P\cdot \d_R)(2R\cdot \ell)^3\ed
	\eea
	The vector $R$ in \eref{set-1-2} is called {\sl auxiliary vector}. Involving the auxiliary vector $R$ not only greatly simplified the tensor structure but also provided a very simple way to establish algebraic relations to solve the reduction problem, as we will review shortly.   
	
	The reduction of \eref{set-1-2} can be denoted as 
	\bea
	I_{n+1}^{(m)}=C^{(m)}_{n+1 \to n+1}I_{n+1}+C^{(m)}_{n+1 \to n+1;\what{i}}I_{n+1;\what{i}}+C^{(m)}_{n+1 \to n+1;\what{ij}}I_{n+1;\what{ij}}+\ldots
	\label{eq:reduction 01}
	\eea
	where $I_{n+1;\what{i_1,\ldots,i_a}}$ represents  the scalar integral got by removing propagators $D_{i_1},D_{i_2},\ldots,D_{i_a}$ from the  integral  \eref{set-1-2}. For example,
	\begin{equation}
		I_{8;\what{0,2,5,6}}=\int {d^D \ell\over i\pi^{D/2}}  {1\over D_1\cdot D_3\cdot D_4\cdot D_7};\ \ \ I_{7;\what{1,3,5}}=\int {d^D \ell\over i\pi^{D/2}}  {1\over D_0\cdot D_2\cdot D_4\cdot D_6}.
	\end{equation}
	For later convenience, in this paper we define
	\bea s_{ij}=K_i\cdot K_j,~~~s_{0i}=R\cdot K_i,~~~f_i=M_0^2-M_i^2+K^2_i~,~~~\label{kinematics}\eea
	and the Gram matrix is given by   $\boldsymbol{G}=[G_{ij}=s_{ij}]$ and its determinant is denoted by $|G|$.
	
	Now we review how to solve reduction coefficients in \eref{eq:reduction 01}.  Without loss of generality, we only review how to calculate  $C^{(m)}_{n+1 \to n+1;\widehat{r+1,r+2,...,n}}$. Other reduction coefficients can be got by a label permutation and  a proper momentum shift.\footnote{More details can be found in \cite{Feng:2021enk,Hu:2021nia}}. Having introduced the $R$,  we can construct the following two types of differential operators
	\bea
	\mathcal{D}_i\equiv K_i\cdot {\d \over \d R},~~i=1,...,n;~~~~ \quad \mathcal{T}\equiv \eta^{\mu\nu}{\d\over \d R^\mu}
	{\d \over \d R^\nu}.\label{def-diffe}
	\eea 
	Acting with these two operators on both sides of \eqref{eq:reduction 01}  and comparing the coefficients of each Master Integral.\footnote{Note that Master Integral contains no $R$, the differential operators will directly act on reduction coefficients. } we get
	\begin{align}
		\mathcal{T} C_{n+1\to n+1;\what{r+1,...,n}}^{(m)}= & 4m(m-1) M_0^2 C_{n+1\to n+1;\what{r+1,...,n}}^{(m-2)}, \label{differential_equation_T}
	\end{align}
	and 
	\begin{align}
		\mathcal{D}_i C_{n+1\to n+1;\what{r+1,...,n}}^{(m)}=& -m  C_{n+1;\what{i}\to n+1;\what{r+1,...,n}}^{(m-1)} + mf_i C_{n+1\to n+1;\what{r+1,...,n}}^{(m-1)}.
		\label{differential_equation_D}
	\end{align}
	where  $C_{n+1;\what{i}\to n+1;\what{r+1,...,n-1}}^{(m-1)}$ is the reduction coefficient of integral $I^{(m-1)}_{n+1;\what{i}}$ to the Master Integral $I_{n+1;\what{r+1,...,n}}$.
	After expanding reduction coefficients according to its tensor structure
	\begin{align}
		C^{(m)}_{n+1 \to n+1;\widehat{r+1,r+2,...,n}}
		=& \sum_{2a_0+\sum_{k=1}^{n}a_k=m}\Bigg\{ c^{(0,1,\cdots,r)}_{a_1,\cdots,a_{n}}(m) (M_0^2)^{a_0+r-n}\prod_{k=0}^{n} s_{0k}^{a_k}\Bigg\} ,\label{coefficient_expansion_1}
	\end{align}
	equations \eref{differential_equation_T} and \eref{differential_equation_D} will lead to
	algebraic recursion relations for  \textit{expansion coefficients} $c^{(0,1,\cdots,r)}_{a_1,\cdots,a_{n}}(m)$ as a rational function of spacetime dimension $D$, rank $m$ and kinematics $(K_i\cdot K_j)$ and $M_i^2$.
	A very nice feature is that these algebraic recursion relations can be solved to give the recursive construction 
	for expansion coefficients as following, 
	%
	%
	\begin{align}
		\boldsymbol{c}^{(0,1,\cdots,r)}(a_1,\cdots,a_n;m)=\boldsymbol{T}^{-1} \widetilde{\boldsymbol{G}}^{-1}  \boldsymbol{O}^{(0,1,\cdots,r)}(a_1,\cdots,a_n;m), \label{inverse_formula}
	\end{align}
	\begin{align}
		c^{(0,1,\cdots,r)}_{\underbrace{0,\cdots,0}_{\text{n\ times}}}(2k)
		={2k-1\over  D+2 k-n-2}\left[(4-\boldsymbol{\a}^T\widetilde{\boldsymbol{G}}^{-1}\boldsymbol{\a})c_{\underbrace{0,\cdots,0}_{\text{n\ times}}}^{(0,1,\cdots,r)}(2k-2)+\boldsymbol{\a}^T\widetilde{\boldsymbol{G}}^{-1}\boldsymbol{c}_{\underbrace{0,\cdots,0}_{\text{n-1\ times}}}^{(0,1,\cdots,r)}(2k-2)\right],\label{Relations}
	\end{align}
	where $\widetilde{\boldsymbol{G}}=[\beta_{ij}={s_{ij}\over M_0^2}]$ is the $n\times n$ dimensionless Gram matrix and  $\boldsymbol{T}=\text{diag}(a_1+1,a_2+1,\cdots,a_n+1)$ is a diagonal matrix. Other vectors are defined as
	\begin{align}
		&\boldsymbol{\a}^T=\left(\frac{f_1}{M_0^2},\frac{f_2}{M_0^2},\cdots,\frac{f_n}{M^2_0}\right) 
		,\\
		&[\boldsymbol{c}^{(0,1,\cdots,r)}(a_1,\cdots,a_n;m)]_i= c^{(0,1,\cdots,r)}_{a_1,a_2,\cdots,a_i+1,\cdots, a_n}(m),
	\end{align}
	and
	\bea
	& & [\boldsymbol{O}^{(0,1,\cdots,r)}(a_1,\cdots,a_n;m)]_i
	= m\alpha_i c^{(0,1,\cdots,r)}_{a_1,\cdots,a_n}(m-1)- m  \delta_{0 a_i}  c^{(0,1,\cdots,r)}_{a_1,\cdots, \widehat{a}_i,\cdots, a_n}(m-1;\widehat{i}) \notag\\
	&&~~~~~~~~~~~~~~~~~~~~~~~~~~~~~~~~~~~~~~-(m+1-\sum_{l=1}^n a_l) c^{(0,1,\cdots,r)}_{a_1,\cdots,a_i-1,\cdots,a_n}(m)\nn
	& & \boldsymbol{c}_{\underbrace{0,\cdots,0}_{\text{n-1\ times}}}^{(0,1,\cdots,r)}(m)=
	\left(0,0,\cdots,0,c_{\underbrace{0,\cdots,0}_{\text{n-1\ times}}}^{(0,1,\cdots,r)}(m;\widehat{r+1}),\cdots,c_{\underbrace{0,\cdots,0}_{\text{n-1\ times}}}^{(0,1,\cdots,r)}(m;\widehat{n-1})\right).\label{define of O}
	\eea
	With the known boundary conditions, one can get all reduction coefficients by applying the recursions \eqref{inverse_formula}, \eref{Relations} iteratively. Above formulas are easy to be implemented into \textsf{Mathematica}
	and one can generate analytic expressions for reduction coefficients for any tensor rank (Exactly because of these benefits involving the auxiliary vector $R$, we call the method as "imporoved PV-reduction
	method"). Knowing the analytic expressions  for reduction coefficients, we can 
	discuss many topics. For example, we can find the reduction for integrals with
	propagators having power higher than one as shown in \cite{Feng:2022uqp}. Another important application is  for the case where some reduction coefficients become divergent,
	as will be discussed in this paper.

	Now we want to point out an important observation from \eqref{inverse_formula}, \eref{Relations}: with ${\boldsymbol{G}}^{-1}$ we will have $|G|$ appearing in denominators 
	of all reduction coefficients. When applying  above PV-reduction method, we have 
	made a hidden assumption that 
	$D\geq n$ and all $K_i$'s in propagators are generic (i.e., linear independent). With this hidden assumption,
	$|G|\neq 0$ and above formulas are well-defined. However, in the practical applications, we do meet the  degeneration of the Gram
	determinant, i.e., $|G|=0$, which means that these $K_i$'s are not linear independent anymore and
	the kinematics lives in a smaller space. For this case, the formulas \eqref{inverse_formula} and  \eref{Relations}
	are ill-defined. Thus, for the completion of the improved PV-reduction method, we must provide an algorithm to solve this
	issue. Briefly, $|G|=0$ means 	
	that the scalar integrals of the highest topology are not in the basis anymore, and they can be reduced to a combination of lower topologies. 
	
	Later we need to use the divergence behavior when $|G|\to 0$. Noticing that 
	$G^{-1}=G^*/|G|$, where $G^*$ is the adjugate matrix of Gram matrix, there is a Gram determinant appearing in the denominator for each iteration, thus one can conclude that the reduction coefficients have the highest divergent part about the determinant of Gram matrix\footnote{In fact, only those reduction coefficients with $m+r\geq n$ are nonzero. Because there are not enough $\ell\cdot R$ in the numerator to cancel $(n-r)$ propagators for $m+r<n$.}
	\begin{equation}
		\begin{aligned}
			&C^{(m)}_{n+1 \to n+1;\what{r+1,...,n}}\sim\left\{\begin{matrix}
				&\frac{1}{|G|^{m}},\ r=n,n-1\\ &\frac{1}{|G|^{m-n+r+1}},\ r<n-1.
			\end{matrix}\right.\\
			&c^{(0,...,n)}_{a_1,...,a_n}\sim |G|^{-\frac{m+\sum_i^n a_i}{2}}
		\end{aligned}
		\label{eq:diverge 01}
	\end{equation}

	\section{Reduction for degenerate Gram Matrix}
	\label{sec:reduction}
	In the previous section, we find that the reduction coefficients diverge when $|G|\to 0$, while $I_{n+1}^{(m)}$ is still well-defined. The naive conflict in \eref{eq:reduction 01}
	indicates that the expansion on the RHS is not proper. In other words, the scalar integrals 
	are not independent of each other anymore. As we will show in this section, these 
	divergences tell us a lot of information about the {degenerating} behavior of the basis.  
	We will use various examples, i.e., integrals of bubble, triangle, box, and pentagon, to demonstrate the strategy to deal with the divergences in our framework.

	\subsection{The reduction of scalar bubble  with degenerate Gram Matrix}
	
	Let us start with the simplest example, i.e., the reduction of the bubble with tensor rank $1$. Our method 
	gives 
	\begin{equation}
		\begin{aligned}
			I_2^{(1)}=&C^{(1)}_{2\to 2}~ I_2+C^{(1)}_{2\to 2;\what{0}}~ I_{2;\what{0}}+C^{(1)}_{2\to 2;\what{1}}~ I_{2;\what{1}}\co
		\end{aligned}
		\label{eq:reduce bubble 01}
	\end{equation}
	where 
	\begin{align}
		\ctoself{1}{2}&=\frac{\left(M_0^2-M_1^2+s_{11}\right) s_{01}}{s_{11}}\co\nn
		\ctonext{1}{2}{1}&=\frac{s_{01}}{s_{11}}\co~~~\ctonext{1}{2}{2}=-\frac{s_{01}}{s_{11}}\ed
		\label{bub-1-1}
	\end{align}
	It is easy to see that with $|G|=k_1\cdot k_1=s_{11}\to 0$, all coefficients in \eref{bub-1-1}
	become divergent. {But $I_2^{(1)}$ is a definitely well-defined integral even 
		with $s_{11}=0$}, so {the divergence on the RHS in \eref{eq:reduce bubble 01} implies} that
	the expansion is not proper. More explicitly, the three scalar integrals $I_{2},I_{2;\what{0}},I_{2;\what{1}}$ are not {linearly} independent anymore, i.e., we should have
	\bea I_{2}= \sum_{i=0}^1 \mathcal{B}_{2\to 2; \what{i}} I_{2;\what{i}}=
	\sum_{i=0}^1 \sum_{a=0}^\infty \mathcal{B}^{(a)}_{2\to 2; \what{i}} s_{11}^a I_{2;\what{i}}\co~~~\label{bub-decom-1}\eea
	where the decomposition coefficient $\mathcal{B}_{2\to 2; \what{i}}$ {is} expanded as  a Taylor series
	of $|G|=s_{11}$.  
	
	Now we show how to use the \eref{bub-decom-1} to make the RHS of \eref{eq:reduce bubble 01} well-defined under the limit $|G|\to 0$. Putting \eref{bub-1-1}
	and \eref{bub-decom-1} to  \eref{eq:reduce bubble 01}, the RHS becomes
	\bea & & \left( \frac{\left(M_0^2-M_1^2+s_{11}\right) s_{01}}{s_{11}}\sum_{a=0}^\infty \mathcal{B}^{(a)}_{2\to 2; \what{0}} s_{11}^a+ \frac{s_{01}}{s_{11}}\right)I_{2;\what{0}}\nn
	& & + \left( \frac{\left(M_0^2-M_1^2+s_{11}\right) s_{01}}{s_{11}}\sum_{a=0}^\infty \mathcal{B}^{(a)}_{2\to 2; \what{1}} s_{11}^a- \frac{s_{01}}{s_{11}}\right)I_{2;\what{1}} .\eea
	{To cancel the divegence, we just need}
	\bea \mathcal{B}^{(0)}_{2\to 2; \what{0}}= {-1\over (M_0^2-M_1^2)}\co~~~\mathcal{B}^{(0)}_{2\to 2; \what{1}}= {1\over (M_0^2-M_1^2)}\ed~~~
	\label{bub-B-0}\eea

	{Although} we have {only} fixed the first coefficient in the above computation, the smoothness of RHS 
	{in} \eref{eq:reduce bubble 01} {does tell} us the information of later coefficients. To get
	{them}, it is natural to consider the reduction of higher tensor rank. 
	For rank $2$, we have 
	\bea
	\ctoself{2}{2}&=&\frac{D \left(M_0^2-M_1^2\right){}^2 s_{01}^2}{(D-1) s_{11}^2}+\frac{D s_{01}^2+2 M_0^2 s_{00}+2 M_1^2 s_{00}}{D-1}\nn 
	&&+\frac{2 M_0^2 \left((D-2) s_{01}^2+M_1^2 s_{00}\right)-2 D M_1^2 s_{01}^2-M_0^4 s_{00}-M_1^4 s_{00}}{(D-1) s_{11}}-\frac{s_{11} s_{00}}{D-1}\co\nn
	\ctonext{2}{2}{1}&=&\frac{D \left(M_0^2-M_1^2\right) s_{01}^2}{(D-1) s_{11}^2}+\frac{(3 D-4) s_{01}^2+M_0^2 \left(-s_{00}\right)+M_1^2 s_{00}}{(D-1) s_{11}}+\frac{s_{00}}{D-1}\co\nn
	\ctonext{2}{2}{2}&=&\frac{D \left(M_1^2-M_0^2\right) s_{01}^2}{(D-1) s_{11}^2}+\frac{-D s_{01}^2+M_0^2 s_{00}-M_1^2 s_{00}}{(D-1) s_{11}}+\frac{s_{00}}{D-1}\ed~~~\label{bub-1-2}
	\eea
	Now the denominators have $s_{11}^2$ in \eref{bub-1-2}. When combining 
	\eref{bub-decom-1} and \eref{bub-1-2}, the smoothness under the limit requires the 
	cancellation of poles ${1\over s_{11}^2}$ and ${1\over s_{11}}$. One can easily check that 
	the result \eref{bub-B-0} will remove the pole ${1\over s_{11}^2}$ automatically, while
	the {cancellation} of the pole ${1\over s_{11}}$ {gives} 
	\bea \mathcal{B}^{(1)}_{2\to 2; \what{0}}&= &  -\frac{D M_0^2-(D-4) M_1^2}{D \left(M_0-M_1\right){}^3 \left(M_0+M_1\right){}^3}\co \nn
	\mathcal{B}^{(1)}_{2\to 2; \what{1}}& = &  -\frac{(D-4) M_0^2-D M_1^2}{D \left(M_0-M_1\right){}^3 \left(M_0+M_1\right){}^3}\ed~~~
	\label{bub-B-1}\eea

	Now the idea is clear. To get the higher-order decomposition coefficients $\mathcal{B}^{(a)}_{2\to 2; \what{i}}$, we just need to impose the smoothness limit of 
	tensor reduction with  higher ranks. In fact, we can 
	{get} expansion coefficients $\mathcal{B}^{(a)}_{2\to 2; \what{i}}$ with $a\leq m$ for bubble {through calculation with tensor rank $(m+1)$}.

	Before ending this subsection, let us point out that the decomposition coefficients given in 
	\eref{bub-B-0} and \eref{bub-B-1} have a nice symmetry between $M_0\leftrightarrow M_1$ {and they become singular under the limit $M_1\to M_0$}. In that limit, $I_{2;\what{0}}$ and $I_{2;\what{1}}$ are not linearly independent of each other, and we should expand
	\bea I_{2;\what{1}} & = & \sum_{a=0}^\infty \mathcal{B}^{(a)}_{(2; \what{1})
		\to (2; \what{0})} (M_1^2-M_0^2)^a I_{2;\what{0}}\ed~~~~\label{tad-decomp-1}\eea
	Putting \eref{tad-decomp-1} into \eref{bub-decom-1} with known $\mathcal{B}^{(a)}_{2\to 2; \what{i}}$ and requiring the smoothness under the limit $M_1\to M_0$, we can {calculate } all the  $\mathcal{B}^{(a)}_{(2; \what{1})
		\to (2; \what{0})}$. For example, with results \eref{bub-B-0} and \eref{bub-B-1},
	we {get}
	\bea
	\mathcal{B}^{(0)}_{(2; \what{1})
		\to (2; \what{0})}=1\co~~~\mathcal{B}^{(1)}_{(2; \what{1})
		\to (2; \what{0})}=\frac{D-2}{2 M_0^2}\co~~~\mathcal{B}^{(2)}_{(2; \what{1})
		\to (2; \what{0})}=\frac{D^2-6 D+8}{8 M_0^4}\ed
	\eea
	\subsection{Algorithm for general case}
	
	Now we are ready to discuss the degeneration of scalar $(n+1)$-gon integral when the Gram determinant $|G|\to 0$. {The Gram matrices corresponding to lower topologies don't degenerate  generally  with $|G|\to 0$}, thus  
	the scalar $(n+1)$-gon integral will become a linear combination of scalar $m$-gon integrals with $m\leq n$, i.e.,
	\begin{equation}
		\begin{aligned}
			&I_{n+1}=\sum_{|\mathbf{b}_j|=1}^{n}\mathcal{B}_{n+1;\what{\mathbf{b}_j}}I_{n+1;\what{\mathbf{b}_j}}= \sum_{|\mathbf{b}_j|=1}^{n}\sum_{a=0}^{\infty}\mathcal{B}^{(a)}_{n+1;\what{\mathbf{b}_j}} |G|^a I_{n+1;\what{\mathbf{b}_j}}\co
		\end{aligned}
		\label{eq:expand n+1-gon 01}
	\end{equation}
	where at the second equation, the Taylor expansion of $|G|$ is given explicitly. 
	To determine these decomposition coefficients $\mathcal{B}^{(a)}_{n+1;\what{\mathbf{b}_j}} $,
	we put \eref{eq:expand n+1-gon 01} to \eref{eq:reduction 01} and get
	\bea I_{n+1}^{(m)}& = & \sum_{|\mathbf{b}_j|=0}^{n}C^{(m)}_{n+1\to n+1;\what{\mathbf{b}_j}}I_{n+1;\what{\mathbf{b}_j}} =C^{(m)}_{n+1\to n+1} I_{n+1}+ \sum_{|\mathbf{b}_j|=1}^{n}C^{(m)}_{n+1\to n+1;\what{\mathbf{b}_j}}I_{n+1;\what{\mathbf{b}_j}}\nn
	& = & \sum_{|\mathbf{b}_j|=1}^{n}\left( C^{(m)}_{n+1\to n+1;\what{\mathbf{b}_j}}+C^{(m)}_{n+1\to n+1} \mathcal{B}_{n+1;\what{\mathbf{b}_j}}\right)I_{n+1;\what{\mathbf{b}_j}}\ed
	~~~\label{gen-dec-1}\eea
	For later convenience, let us write  
	\bea \mathcal{F}^{(m)}_{n+1\to n+1;\what{\mathbf{b}_j}}= \left( C^{(m)}_{n+1\to n+1;\what{\mathbf{b}_j}}+C^{(m)}_{n+1\to n+1} \mathcal{B}_{n+1;\what{\mathbf{b}_j}}\right)\ed~~~\label{gen-dec-2}\eea
	The smoothness of $I_{n+1}^{(m)}$ under the limit $|G|\to 0$ means that  $\mathcal{F}^{(m)}_{n+1\to n+1;\what{\mathbf{b}_j}}$ is a Taylor series of $|G|$. Using 
	this condition, we can {get} all the
	decomposition coefficients. 
	
	Now we show how to do it {more directly} instead of solving linear equations.  
	Since $C^{(m)}_{n+1\to n+1;\what{\mathbf{b}_j}}$ is a Laurent  series about the Gram determinant $|G|$, for example, 
	\bea
	C^{(m)}_{n+1\to n+1}=\sum_{i=-m}^{\infty}C^{(m;i)}_{n+1\to n+1}|G|^i\co~~~\label{C-h}
	\eea 
	{using \eref{gen-dec-2}, we can write 
		\begin{equation}
			\begin{aligned}
				\mathcal{B}_{n+1;\what{\mathbf{b}_j}}=&\frac{\mathcal{F}^{(m)}_{n+1\to n+1;\what{\mathbf{b}_j}}-C^{(m)}_{n+1\to n+1;\what{\mathbf{b}_j}}}{C^{(m)}_{n+1\to n+1}}
				=-\frac{C^{(m)}_{n+1\to n+1;\what{\mathbf{b}_j}}}{C^{(m)}_{n+1\to n+1}}+\frac{\mathcal{F}^{(m)}_{n+1\to n+1;\what{\mathbf{b}_j}}}{C^{(m)}_{n+1\to n+1}}\ed
			\end{aligned}
			\label{Bubi-Rationfunction}
		\end{equation}
		In \eqref{Bubi-Rationfunction}, $\mathcal{F}^{(m)}_{n+1\to n+1;\what{\mathbf{b}_j}}$ is regular and $C^{(m)}_{n+1\to n+1}\sim |G|^{-m}$, so we just need to
		keep the first term in the numerator of \eref{Bubi-Rationfunction} to get the coefficients $\mathcal{B}^{(k<m)}_{n+1;\what{\mathbf{b}_j}}$, i.e.,}
	\bea
	\mathcal{B}_{n+1;\what{\mathbf{b}_j}}= -{\frac{C^{(m)}_{n+1\to n+1;\what{\mathbf{b}_j}}}{C^{(m)}_{n+1\to n+1}}}+ \mathcal{O}(|G|^m)\ed~~~\label{B-m}
	\eea
	Using \eref{B-m}, we can get
	\bea \mathcal{F}^{(m)}_{n+1\to n+1;\what{\mathbf{b}_j}}&=& \left( C^{(m)}_{n+1\to n+1;\what{\mathbf{b}_j}}+\left(-{\frac{C^{(m')}_{n+1\to n+1;\what{\mathbf{b}_j}}}{C^{(m')}_{n+1\to n+1}}}+ \mathcal{O}(|G|^{m'})\right)C^{(m)}_{n+1\to n+1} \right)\nn
	&=&{\frac{C^{(m)}_{n+1\to n+1;\what{\mathbf{b}_j}}C^{(m')}_{n+1\to n+1} -C^{(m')}_{n+1\to n+1;\what{\mathbf{b}_j}}C^{(m)}_{n+1\to n+1}}{C^{(m')}_{n+1\to n+1}}}+  \mathcal{O}(|G|^{m'-m})\ed
	~~~\label{F-m}\eea
	{Here we use the fact $C^{(m)}_{n+1\to n+1}\sim |G|^{-m}$. To get the accurate expansion of $\mathcal{F}^{(m)}_{n+1\to n+1;\what{\mathbf{b}_j}}$ up to $k$-th order of 
		$|G|$, we set 
		$m'-m = k+1$, then we can write }
	\bea \mathcal{F}^{(m)}_{n+1\to n+1;\what{\mathbf{b}_j}}= \frac{C^{(m)}_{n+1\to n+1;\what{\mathbf{b}_j}}C^{(m+k+1)}_{n+1\to n+1} -C^{(m+k+1)}_{n+1\to n+1;\what{\mathbf{b}_j}}C^{(m)}_{n+1\to n+1}}{C^{(m+k+1)}_{n+1\to n+1}}+ {\cal O}(|G|^{k+1})\ed~~~\label{F-m-1}\eea
	{If we take $k=0$ in \eref{F-m-1}, it gives } 
	\bea F^{(m)}_{n+1\to n+1;\what{\mathbf{b}_j}}=\lim_{|G|\to 0} \left( \frac{C^{(m)}_{n+1\to n+1;\what{\mathbf{b}_j}}C^{(m+1)}_{n+1\to n+1} -C^{(m+1)}_{n+1\to n+1;\what{\mathbf{b}_j}}C^{(m)}_{n+1\to n+1}}{C^{(m+1)}_{n+1\to n+1}} \right)\co~~~\label{F-m-2}\eea
	{which gives the reduction coefficients for $I_{n+1}^{(m)}$ when $|G|=0$.} The same result \eref{F-m-2}
	has been obtained {by methods in projective space} in \cite{Feng:2022rwj}. For the expansion \eref{eq:expand n+1-gon 01},
	if we care about only the zeroth order, the result is also well known, which is given by \cite{Denner:1991kt,Melrose:1965kb,Ossola_2007} (see Eq(5.8) in \cite{Ossola_2007})
	\begin{equation}
		\left|
		\begin{array}{ccccc}
			I_{n+1}    & -I_{n+1,\what{0}}    & -I_{n+1,\what{1}}    & \ldots & -I_{n+1,\what{n}} \\
			1      & Y_{0\,0}   & Y_{0\,1}   & \ldots & Y_{0\,n} \\
			1      & Y_{1\,0}   & Y_{1\,1}   & \ldots & Y_{1\,n} \\
			\vdots & \vdots     & \vdots     & \vdots & \vdots \\
			1      & Y_{n\,0} & Y_{n\,1} & \ldots & Y_{n\,n}
		\end{array}
		\right | =0\,,
		~~~\label{det=0}
	\end{equation}
	with 
	\bea Y_{ij}=M_i^2+N_j^2-(K_i-K_j)^2, \,\,\, i,j=0,\ldots,n. \,\,\,
	\eea
	A good feature of the result \eref{B-m} is that we can get the higher-order expansion, which may be helpful for a better numerical evaluation around {the point $|G|=0$}. 
	
	Having established the general formula in \eref{B-m} and \eref{F-m-2}, let us redo the 
	example of the bubble in the previous subsection:
	\begin{itemize}
		
		\item Up to the order $s_{11}^0$: we have 
		\bea
		\mathcal{B}_{2;\what{1}} & = & -{\frac{C^{(1)}_{2\to 2;\what{1}}}{C^{(1)}_{2\to 2}}}=\frac{1}{M_0^2-M_1^2}+\mathcal{O}\left(|G|\right)\co\nn
		\mathcal{B}_{2;\what{0}} & = & -{\frac{C^{(1)}_{2\to 2;\what{0}}}{C^{(1)}_{2\to 2}}}=\frac{1}{M_1^2-M_0^2}+\mathcal{O}\left(|G|\right)\ed
		\eea
		after using the result \eref{bub-1-1}. 
		
		\item Up to the order $s_{11}^1$: we have 
		\bea
		\mathcal{B}_{2;\what{1}}&= & -{\frac{C^{(2)}_{2\to 2;\what{1}}}{C^{(2)}_{2\to 2}}}= \frac{1}{M_0^2-M_1^2}+\frac{ \Big(D M_1^2-(D-4) M_0^2\Big)|G|}{D \left(M_0^2-M_1^2\right){}^3}+\mathcal{O}\left(|G|^2\right)\co\nn
		\mathcal{B}_{2;\what{0}} & = &-{\frac{C^{(2)}_{2\to 2;\what{0}}}{C^{(2)}_{2\to 2}}}=\frac{1}{M_1^2-M_0^2}+\frac{ \Big((D-4) M_1^2-D M_0^2\Big)|G|}{D \left(M_0^2-M_1^2\right){}^3}+\mathcal{O}\left(|G|^2\right)\ed
		\eea
		after using the result \eref{bub-1-2}. 
		
		\item 	Up to the order $s_{11}^2$: we have 
		\bea
		\mathcal{B}_{2;\what{1}}&= & -{\frac{C^{(3)}_{2\to 2;\what{1}}}{C^{(3)}_{2\to 2}}}= \frac{1}{M_0^2-M_1^2}+\frac{\Big(D M_1^2-(D-4) M_0^2\Big) |G|}{D \left(M_0^2-M_1^2\right){}^3}\nn
		&&+\frac{ \Big(\left(D^2-10 D+24\right) M_0^4-2 \left(D^2-4 D-12\right) M_1^2 M_0^2+D (D+2) M_1^4\Big)|G|^2}{D (D+2) \left(M_0^2-M_1^2\right){}^5}\nn
		&&+\mathcal{O}\left(|G|^3\right)\co\nn
		\mathcal{B}_{2;\what{0}} & = & -{\frac{C^{(3)}_{2\to 2;\what{0}}}{C^{(3)}_{2\to 2}}}=\frac{1}{M_1^2-M_0^2}+\frac{\Big((D-4) M_1^2-D M_0^2\Big)|G|}{D \left(M_0^2-M_1^2\right){}^3}\nn
		&&-\frac{ \Big(\left(D^2-10 D+24\right) M_1^4-2 \left(D^2-4 D-12\right) M_1^2 M_0^2+D (D+2) M_0^4\Big)|G|^2}{D (D+2) \left(M_0^2-M_1^2\right){}^5}\nn
		&&+\mathcal{O}\left(|G|^3\right)\ed
		\eea
		after using the result in \cite{Hu:2021nia} (see Eq(4.11) to Eq(4.16)). 	
		
	\end{itemize}

	\subsection{The degenerate triangle}
	
	In this subsection, we will present some results for the degenerate triangle. First, we parameterize the Gram determinant as\footnote{One should be very careful about the kinematics. For example, in the calculation of tensor triangle, we cannot treat $s_{11},s_{12},s_{22}$ and $|G|$ as four independent Lorentz invariant combinations. Instead, we should replace, for example, every $s_{22}$ with $(|G|+s^2_{12})/s_{11}$.}
	\bea G_{\text{tri}}=s_{11} s_{22}-s_{12}^2=|G|\co
	\eea
	where $s_{ij}\equiv K_i\cdot K_j$ and the parameter $|G|$ controls the speed of degeneration. We expand the Master Integral $I_3$ as
	\bea 
	I_{3}&=&\sum_{i_1=0}^{2}\mathcal{B}_{3;\what{i_1}}I_{3;\what{i_1}}+\sum_{0\le i_1< i_2}^{2}\mathcal{B}_{3;\what{i_1i_2}}I_{3;\what{i_1i_2}}\ed
	\eea

	{We will calculate several typical  $\mathcal{B}_{3;\what{\mathbf{b}_j}}$ by \eref{B-m}, while others can be obtained
		by proper permutation.}
	\begin{itemize}
		\item $\mathcal{B}_{3;\what{2}}$: Using the tensor rank {1} we get
		\bea
		\mathcal{B}_{3;\what{2}}=\frac{s_{11}}{M_0^2 \left(s_{11}-s_{12}\right)-M_2^2 s_{11}+s_{12} \left(	M_1^2-s_{11}+s_{12}\right)}+\mathcal{O}\left(|G|\right)\co~~~\label{tri-2-a}
		\eea
		while using the tensor rank 2 we get
		\bea
		\mathcal{B}_{3;\what{2}}&=&\frac{s_{11}}{M_0^2 \left(s_{11}-s_{12}\right)-M_2^2 s_{11}+s_{12} \left(M_1^2-s_{11}+s_{12}\right)}\nn
		&&-\frac{s_{11} |G| }{(D-1) \left(M_0^2 \left(s_{11}-s_{12}\right)-M_2^2 s_{11}+s_{12} \left(M_1^2-s_{11}+s_{12}\right)\right){}^3}\times\nn
		&&\Big(s_{11} \left(D M_0^2-D M_2^2-3 M_0^2-2 M_1^2+M_2^2\right)+(D-1) s_{12}^2+M_1^4+s_{11}^2+\nn
		&&s_{12} \left(-D M_0^2+D M_1^2+M_0^2-M_1^2\right)+M_0^4-2 M_1^2 M_0^2+(1-D) s_{11} s_{12}\Big)\nn
		&&+\mathcal{O}\left(|G|^2\right)\ed~~~~~\label{tri-2-b}\eea
		One can see that for zeroth order, results in \eref{tri-2-a} and \eref{tri-2-b} are same. Similar results hold for other
		coefficients. 
		
		\item $\mathcal{B}_{3;\what{0}}$: Using the tensor rank 1 we get
		\bea
		\mathcal{B}_{3;\what{0}}=\frac{s_{12}-s_{11}}{M_0^2 \left(s_{11}-s_{12}\right)-M_2^2 s_{11}+s_{12} \left(M_1^2-s_{11}+s_{12}\right)}+\mathcal{O}\left(|G|\right)\co
		\eea
		while using the tensor rank 2 we get
		\bea
		\mathcal{B}_{3;\what{0}}=&&\frac{s_{12}-s_{11}}{M_0^2 \left(s_{11}-s_{12}\right)-M_2^2 s_{11}+s_{12} \left(M_1^2-s_{11}+s_{12}\right)}\nn
		&&+\frac{|G| }{(D-1) \left(s_{11}-s_{12}\right) \left(M_0^2 \left(s_{12}-s_{11}\right)+M_2^2 s_{11}-s_{12} \left(M_1^2-s_{11}+s_{12}\right)\right){}^3}\times\nn
		&&\Big(s_{11}^2 \big(D M_0^2 M_1^2-D M_2^2 M_1^2+D M_2^4-D M_0^2 M_2^2-M_1^4-M_0^2 M_1^2+3 M_2^2 M_1^2-\nn
		&&2 M_2^4+M_0^2 M_2^2\big)+s_{12} s_{11}^2 \left(-D M_1^2+D M_2^2-3 M_1^2-5 M_2^2\right)+s_{12}^2 s_{11} \big(D M_1^2-\nn
		&&D M_2^2+M_1^2+3 M_2^2\big)+s_{12} s_{11} \big(D M_1^4-D M_0^2 M_1^2-D M_2^2 M_1^2+D M_0^2 M_2^2-\nn
		&&M_1^4+M_0^2 M_1^2+M_2^2 M_1^2-M_0^2 M_2^2\big)+\left(2 M_1^2+2 M_2^2\right) s_{11}^3-s_{11}^4+4 s_{12} s_{11}^3-\nn
		&&6 s_{12}^2 s_{11}^2+4 s_{12}^3 s_{11}-s_{12}^4\Big)+\mathcal{O}\left(|G|^2\right)\ed
		\eea

		\item $\mathcal{B}_{3;\what{12}}$: Using the tensor rank 1 we get
		\bea
		\mathcal{B}_{3;\what{12}}=0+\mathcal{O}\left(|G|\right)\co
		\eea
		while using the tensor rank 2 we get
		\bea
		\mathcal{B}_{3;\what{12}}=\frac{(D-2) s_{11} |G|}{(D-1) s_{12} \left(M_0^2 \left(s_{11}-s_{12}\right)-M_2^2 s_{11}+s_{12} \left(M_1^2-s_{11}+s_{12}\right)\right){}^2}+\mathcal{O}\left(|G|^2\right)\ed\nn~~~~~
		\eea
		Finally using the tensor rank 3 we get
		\bea
		\mathcal{B}_{3;\what{12}}&=&\frac{(D-2) s_{11} |G|}{(D-1) s_{12} \left(M_0^2 \left(s_{11}-s_{12}\right)-M_2^2 s_{11}+s_{12} \left(M_1^2-s_{11}+s_{12}\right)\right){}^2}\nn
		&&-\frac{(D-2) s_{11} |G|^2 }{\left(D^2-1\right) s_{12}^3 \left(M_0^2 \left(s_{11}-s_{12}\right)-M_2^2 s_{11}+s_{12} \left(M_1^2-s_{11}+s_{12}\right)\right){}^4}\times\nn
		&&\Big(s_{12}^3 \left(-3 D M_0^2+3 D M_1^2-2 M_0^2+2 M_1^2\right)+s_{12}^2 \big(D M_0^4-2 D M_1^2 M_0^2+\nn
		&&D M_1^4+2 M_0^4-4 M_1^2 M_0^2+2 M_1^4\big)+s_{11} s_{12}^2 \big(4 D M_0^2-2 D M_1^2-2 D M_2^2-\nn
		&&4 M_0^2-4 M_1^2-4 M_2^2\big)+s_{11}^2 s_{12} \left(D M_2^2-D M_0^2\right)+s_{11} s_{12} \big(-D M_0^4+\nn
		&&D M_1^2 M_0^2+D M_2^2 M_0^2-D M_1^2 M_2^2\big)+(2 D+3) s_{12}^4+(-3 D-2) s_{11} s_{12}^3\nn
		&&+(D+2) s_{11}^2 s_{12}^2+\left(M_0^4-2 M_2^2 M_0^2+M_2^4\right) s_{11}^2\Big)
		+\mathcal{O}\left(|G|^3\right)\ed~~~~~~~
		\eea
		\item $\mathcal{B}_{3;\what{02}}$: Using the tensor rank 1 we get
		\bea
		\mathcal{B}_{3;\what{02}}=0+\mathcal{O}\left(|G|\right)\co
		\eea
		while using the tensor rank 2 we get
		\bea
		\mathcal{B}_{3;\what{02}}=&&\frac{(D-2) s_{11} |G|}{(D-1) \left(s_{11}-s_{12}\right) \left(M_0^2 \left(s_{12}-s_{11}\right)+M_2^2 s_{11}-s_{12} \left(M_1^2-s_{11}+s_{12}\right)\right){}^2}\nn
		&&+\mathcal{O}\left(|G|^2\right)\ed
		\eea
		Finally using the tensor rank 3 we get
		\begin{align}
			\mathcal{B}_{3;\what{02}}=&\frac{(D-2) s_{11} |G|}{(D-1) \left(s_{11}-s_{12}\right) \left(M_0^2 \left(s_{12}-s_{11}\right)+M_2^2 s_{11}-s_{12} \left(M_1^2-s_{11}+s_{12}\right)\right){}^2}\nn
			&-\frac{(D-2) s_{11} |G|^2 }{\left(D^2-1\right) \left(s_{11}-s_{12}\right){}^3 \left(M_0^2 \left(s_{12}-s_{11}\right)+M_2^2 s_{11}-s_{12} \left(M_1^2-s_{11}+s_{12}\right)\right){}^4}\times\nn
			&\Big(s_{12} s_{11} \big(-2 D M_0^4+3 D M_1^2 M_0^2+D M_2^2 M_0^2-D M_1^4-D M_1^2 M_2^2-4 M_0^4+\nn
			&8 M_1^2 M_0^2-4 M_1^4\big)+s_{11}^3 \left(D M_0^2-D M_2^2-2 M_0^2-6 M_1^2-4 M_2^2\right)+s_{11}^2 \big(D M_0^4-\nn
			&D M_1^2 M_0^2-D M_2^2 M_0^2+D M_1^2 M_2^2+2 M_0^4-4 M_1^2 M_0^2+3 M_1^4+M_2^4-2 M_1^2 M_2^2\big)\nn
			&+s_{12} s_{11}^2 \left(-5 D M_0^2+2 D M_1^2+3 D M_2^2+2 M_0^2+14 M_1^2+8 M_2^2\right)+s_{12}^3 \big(2 M_1^2-\nn
			&3 D M_0^2+3 D M_1^2-2 M_0^2\big)+s_{12}^2 s_{11} \big(7 D M_0^2-5 D M_1^2-2 D M_2^2+2 M_0^2-4 M_2^2-\nn
			&10 M_1^2\big)-(D+10) s_{12} s_{11}^3+(4 D+14) s_{12}^2 s_{11}^2-(5 D+10) s_{12}^3 s_{11}+3 s_{11}^4+\nn
			&(2 D+3) s_{12}^4+s_{12}^2 \left(D M_0^4-2 D M_1^2 M_0^2+D M_1^4+2 M_0^4-4 M_1^2 M_0^2+2 M_1^4\right)\Big)\nn
			&+\mathcal{O}\left(|G|^3\right)\ed	
		\end{align}
	\end{itemize}

	{Note that the expansion of $\mathcal{B}_{n+1;\what{\mathbf{b}_j}}$ is independent of $R$ although the numerator and the denominator depend on $R$ on the RHS of \eref{B-m}. And it's also independent of the tensor rank $m$ up to the order $|G|^k$ as long as $m> k$. The above results confirm these nontrivial points, and we will give a proof in the later section. }

	The  reduction coefficients $F^{(m)}_{3\to 3;\what{\mathbf{b}_j}}$ for $|G|=0$ (see \eref{F-m-2})  are given in following:
	\begin{itemize}
		\item $m=0$:
		\begin{subequations}
			\allowdisplaybreaks
			\bea
			F^{(0)}_{3\to 3;\what{0}}&=&\mathcal{B}^{(0)}_{3\to 3;\what{0}}=-\left(\frac{C^{(1)}_{3\to 3;\what{0}}}{C^{(1)}_{3\to 3}}\right)\Bigg\vert_{\abs{G}=0}\nn
			&=&\frac{s_{12}-s_{11}}{M_1^2 \left(s_{11}-s_{12}\right)-M_3^2 s_{11}+s_{12} \left(M_2^2-s_{11}+s_{12}\right)}\co
			\eea
			\bea
			F^{(0)}_{3\to 3;\what{1}}=-\frac{s_{12}}{M_1^2 \left(s_{11}-s_{12}\right)-M_3^2 s_{11}+s_{12} \left(M_2^2-s_{11}+s_{12}\right)}\co
			\eea
			\bea
			F^{(0)}_{3\to 3;\what{2}}=\frac{s_{11}}{M_1^2 \left(s_{11}-s_{12}\right)-M_3^2 s_{11}+s_{12} \left(M_2^2-s_{11}+s_{12}\right)}\co
			\eea
			\bea
			F^{(0)}_{3\to 3;\what{12}}=F^{(0)}_{3\to 3;\what{01}}=F^{(0)}_{3\to 3;\what{02}}=0\ed
			\eea
		\end{subequations}

		\item $m=1$:
		\begin{subequations}
			\allowdisplaybreaks
			\begin{align}
				F^{(1)}_{3\to3;\what{0}}=&\frac{1}{(D-1) s_{11} \left(s_{11}-s_{12}\right) \left(\left(s_{12}-s_{11}\right) M_1^2+M_3^2 s_{11}-s_{12} \left(M_2^2-s_{11}+s_{12}\right)\right){}^2}\times\nn
				&\Big(s_{02} s_{11}^5+\left(-D M_1^2+M_1^2+D M_3^2-M_3^2\right) s_{01} s_{11}^4+(D-2) s_{12} s_{01} s_{11}^4+\nn
				&\left(-D M_1^2+M_1^2-2 M_2^2+D M_3^2-3 M_3^2\right) s_{02} s_{11}^4+(D-5) s_{12} s_{02} s_{11}^4+\nn
				&(7-3 D) s_{12}^2 s_{01} s_{11}^3+\big(D M_3^4-M_3^4-D M_1^2 M_3^2+M_1^2 M_3^2-D M_2^2 M_3^2+\nn
				&M_2^2 M_3^2+D M_1^2 M_2^2-M_1^2 M_2^2\big) s_{01} s_{11}^3+\big(3 D M_1^2-3 M_1^2-2 D M_2^2+\nn
				&4 M_2^2-D M_3^2+3 M_3^2\big) s_{12} s_{01} s_{11}^3+(9-3 D) s_{12}^2 s_{02} s_{11}^3+(5-D) s_{12}^4 s_{01} s_{11}\nn
				&+\left(3 D M_1^2-3 M_1^2+4 M_2^2-3 D M_3^2+7 M_3^2\right) s_{12} s_{02} s_{11}^3+\big(-3 D M_1^2+\nn
				&3 M_1^2+3 D M_2^2-7 M_2^2-4 M_3^2\big) s_{12}^2 s_{01} s_{11}^2+\big(D M_2^4-2 M_2^4-D M_1^2 M_2^2+\nn
				&M_1^2 M_2^2-D M_3^2 M_2^2+3 M_3^2 M_2^2-M_3^4+D M_1^2 M_3^2-M_1^2 M_3^2\big) s_{12} s_{01} s_{11}^2\nn
				&+(3 D-7) s_{12}^3 s_{02} s_{11}^2+\big(-3 D M_1^2+3 M_1^2+D M_2^2-3 M_2^2+2 D M_3^2\nn
				&-4 M_3^2\big) s_{12}^2 s_{02} s_{11}^2+\big(-D M_2^4+M_2^4+D M_1^2 M_2^2-M_1^2 M_2^2+D M_3^2 M_2^2\nn
				&-M_3^2 M_2^2-D M_1^2 M_3^2+M_1^2 M_3^2\big) s_{12} s_{02} s_{11}^2+\big(M_2^4-D M_1^2 M_2^2+M_1^2 M_2^2+\nn
				&D M_3^2 M_2^2-3 M_3^2 M_2^2-D M_3^4+2 M_3^4+D M_1^2 M_3^2-M_1^2 M_3^2\big) s_{02} s_{11}^3\nn
				&+\big(D M_1^2-M_1^2-D M_2^2+3 M_2^2+2 M_3^2\big) s_{12}^3 s_{01} s_{11}+(2-D) s_{12}^4 s_{02} s_{11}\nn
				&+\big(D M_1^2-M_1^2-D M_2^2+M_2^2\big) s_{12}^3 s_{02} s_{11}-s_{12}^5 s_{01}+(3 D-9) s_{12}^3 s_{01} s_{11}^2\Big)\co\nn
			\end{align}
			\bea
			F^{(1)}_{3\to3;\what{1}}&=&\frac{1}{(D-1) s_{11} s_{12} \left(M_1^2 \left(s_{12}-s_{11}\right)+M_3^2 s_{11}-s_{12} \left(M_2^2-s_{11}+s_{12}\right)\right){}^2}\times\nn
			&&\Big(s_{11} s_{12}^3 \left(D M_1^2-D M_2^2-M_1^2+M_2^2\right) s_{02}+s_{11}^2 s_{12}^2 \big(-2 D M_1^2+2 D M_3^2-\nn
			&&4 M_3^2\big) s_{02}+s_{11}^3 s_{12} \left(D M_1^2-D M_3^2-M_1^2+M_3^2\right) s_{02}+(2-D) s_{11} s_{12}^4 s_{02}\nn
			&&+s_{11}^2 s_{12} \big(D M_1^4-D M_2^2 M_1^2-D M_3^2 M_1^2+D M_2^2 M_3^2-M_1^4+M_2^2 M_1^2+\nn
			&&M_3^2 M_1^2-M_2^2 M_3^2\big) s_{02}+(D-1) s_{11}^2 s_{12}^3 s_{02}+\left(2 M_1^2+2 M_3^2\right) s_{11} s_{12}^3 s_{01}\nn
			&&+s_{11}^3 \left(-D M_1^4+2 D M_3^2 M_1^2-D M_3^4+2 M_1^4-4 M_3^2 M_1^2+2 M_3^4\right) s_{02}\nn
			&&+\left(-M_1^4+2 M_3^2 M_1^2-M_3^4\right) s_{11}^2 s_{12} s_{01}-s_{12}^5 s_{01}\Big)\co
			\eea
			\bea
			F^{(1)}_{3\to3;\what{2}}&=&\frac{1}{(D-1) \left(M_1^2 \left(s_{11}-s_{12}\right)-M_3^2 s_{11}+s_{12} \left(M_2^2-s_{11}+s_{12}\right)\right){}^2}\times\nn
			&&\Big(s_{11}^2 \left(D M_1^2-D M_3^2-M_1^2+M_3^2\right) s_{01}+s_{11} \big(D M_1^4-D M_2^2 M_1^2-\nn
			&&D M_3^2 M_1^2+D M_2^2 M_3^2-M_1^4+M_2^2 M_1^2+M_3^2 M_1^2-M_2^2 M_3^2\big) s_{01}\nn
			&&+s_{12} s_{11} \left(-2 D M_1^2+2 D M_2^2-4 M_2^2\right) s_{01}+s_{12}^2 \big(D M_1^2-D M_2^2-\nn
			&&M_1^2+M_2^2\big) s_{01}+s_{12} \big(-D M_1^4+2 D M_2^2 M_1^2-D M_2^4+2 M_1^4\nn
			&&-4 M_2^2 M_1^2+2 M_2^4\big) s_{01}+(2-D) s_{12} s_{11}^2 s_{01}+(D-1) s_{12}^2 s_{11} s_{01}\nn
			&&+\left(2 M_1^2+2 M_2^2\right) s_{11}^2 s_{02}+\left(-M_1^4+2 M_2^2 M_1^2-M_2^4\right) s_{11} s_{02}\nn
			&&-s_{11}^3 s_{02}\Big)\co
			\eea
			\bea
			F^{(1)}_{3\to3;\what{12}}&=&-\frac{(D-2) \left(s_{12} s_{01}-s_{11} s_{02}\right)}{(D-1) s_{12} \left(M_1^2 \left(s_{11}-s_{12}\right)-M_3^2 s_{11}+s_{12} \left(M_2^2-s_{11}+s_{12}\right)\right)}\co\nn
			F^{(1)}_{3\to3;\what{01}}&=&\frac{(D-2) s_{11} \left(s_{11} s_{02}-s_{12} s_{01}\right)}{(D-1) \left(s_{11}-s_{12}\right) s_{12} \left(M_1^2 \left(s_{12}-s_{11}\right)+M_3^2 s_{11}-s_{12} \left(M_2^2-s_{11}+s_{12}\right)\right)}\co\nn
			F^{(1)}_{3\to3;\what{02}}&=&\frac{(D-2) \left(s_{12} s_{01}-s_{11} s_{02}\right)}{(D-1) \left(s_{11}-s_{12}\right) \left(M_1^2 \left(s_{12}-s_{11}\right)+M_3^2 s_{11}-s_{12} \left(M_2^2-s_{11}+s_{12}\right)\right)}\ed\nn
			\eea
		\end{subequations}
		One interesting point is that expressions of the first three coefficients are complicated while the last three are
		relatively simple. The reason is that $I_{n+1}$ is reduced to combinations of $I_{n}$ only (see \eref{det=0})
		at zeroth order, thus coefficients of $I_{m\leq n-1}$ are not affected. 
		
	\end{itemize}

	In this section, we present examples up to the triangle. A complete result of $\cal{B}$ is listed in the attachment \textsf{Mathematica} file. For box and pentagon, we will give some numerical results in Appendix \ref{appdix:A}.

	\subsection{Two-loop Example: Sunset}
	\label{subsec:sunset}
	Our idea to deal with degenerate Gram determinant is simple and it should be applicable to higher loops. 
	To demonstrate this point, in this subsection we consider the simplest two-loop integrals, i.e., the integrals with sunset topology. 
	Involving two auxiliary vectors $R_1,R_2$ we consider the tensor reduction of following integrals
	\bea
	I_{a_1,a_2,a_3}^{(r_1,r_2)}\equiv \int {d^D\ell_1\over i\pi^{D/2}}{d^D\ell_2\over i\pi^{D/2}} {(2\ell_1\cdot R_1)^{r_1}(2\ell_2\cdot R_2)^{r_2}\over D_1^{a_1}D_2^{a_2}D_3^{a_3}}\co\label{tensor-sunset-def}
	\eea
	where the propagators are
	\bea 
	D_1\equiv \ell_1^2-M_1^2\co~~~D_2\equiv \ell_2^2-M_2^2\co~~~D_3\equiv(\ell_1+\ell_2-K)^2-M_3^2\ed~~~\label{3-D}
	\eea
	It is found that the integrals \eref{tensor-sunset-def} can be reduced to seven master integrals\footnote{For our
		purpose, we consider only the case $a_i=1$.}, i.e., 
	%
	\bea
	I_{1,1,1}^{(r_1,r_2)}={\bf C}^{(r_1,r_2)}{\bf J}=\sum_{i_1,i_2,j} s_{01}^{i_1}s_{0'1}^{i_2}s_{00'}^js_{00}^{r_1-i_1-j\over 2}s_{0'0'}^{r_2-i_2-j\over 2}{\vec{\a}}^{(r_1,r_2)}_{i_1,i_2,j}~{\bf J}\ed \label{coeff-exp-sub}
	\eea
	where the seven master integrals are (we have written ${\bf C}$, ${\bf J}$ to emphasize they are 
	vector)
	\begin{align}
		\mathbf{J}&=\Bigg\{\int {1\over D_1D_2D_3},\int {(2\ell_1 \cdot K) \over D_1D_2D_3},\int {(2\ell_2 \cdot K) \over D_1D_2D_3},\int {(2\ell_1 \cdot K)(2\ell_2 \cdot K) \over D_1D_2D_3}\nn
		&\hspace{40pt}\int {1\over D_2D_3},\int {1\over D_1D_3},\int {1\over D_1D_2}\Bigg\}\co~~~\label{J-basis}
	\end{align}
	and  kinematic variables are 
	\begin{align}
		s_{01}=R_1\cdot K\co~ s_{0'1}=R_2\cdot K\co~ s_{11}=K^2\co~s_{00}=R_1^2\co~s_{0'0'}=R_2^2\co~s_{00'}=R_1\cdot R_2\ed
	\end{align}
	All the reduction coefficients could be found in \cite{Feng:2022iuc}, where we have generalized the improved PV-reduction method to the two-loop sunset topology.  For example:
	\bea
	\mathbf{C}^{(1,0)}&=&\left\{0,\frac{s_{01}}{s_{11}},0,0,0,0,0\right\}\co\nn  \mathbf{C}^{(0,1)}&=&\left\{0,0,\frac{s_{0'1}}{s_{11}},0,0,0,0\right\}\ed \label{sunsetCr1}
	\eea
	
	To study the reduction for degenerate Gram determinant ($s_{11}\to 0$), we follow the same logic, i.e., the 
	basis with the highest topology will be decomposed to the linear combinations of other integrals in the basis. However, 
	there is something different between two-loop and one-loop integrals. For one-loop integrals, there
	is only one master integral with the highest topology in the basis. For two-loop integrals, in general, there are more than one highest-topology master integrals.  For example, for the sunset topology, there are four master integrals $J_1,J_2, J_3, J_4$ as
	given in  \eref{J-basis}. Now we face the problem: whether just one  or several master integrals of them should be degenerate in the limit?  In fact, from the tensor reduction results given in \eref{sunsetCr1}, we can see 
	$J_2$ and $J_3$ should be degenerate. There is also a brute-force algorithm to determine the number of degenerate master integrals. Starting with just one degenerate master integral, we solve the decomposition coefficients using the idea
	given in the last subsection for various choices of tensor rank. If it is not the right degenerate pattern, we will
	find that it is impossible to cancel all divergences coming from $s_{11}\to 0$. Then we take two master integrals to be
	degenerate, repeat the same computations for several tensor structures and check the consistency. Interacting the procedure until reaching the point that all divergences have been canceled consistently, we get the correct
	degenerate pattern. 
	
	For our sunset topology, because there are five independent Lorentz invariant contractions $R_i\cdot R_j,R_i\cdot K$ comparing to  two $R^2,R\cdot K$ for one-loop bubble, there are three master integrals degenerated under the limit, 
	which can be taken as  $J_2,J_3,J_4$, i.e., we have the expansion 
	\bea
	J_i=\sum_{k=0}^\infty {\cal B}_{i\to j}^{(k)}s_{11}^kJ_j\co~~~i\in\text{deI}\co ~j\in\text{reI}\co
	\eea 
	where $\text{deI}=\{2,3,4\}$ denotes the degenerate master integrals and $\text{reI}=\{1,5,6,7\}$ denotes the remaining
	master integrals. Expansion coefficients ${\cal B}_{i\to j}^{(k)}$ are determined by requiring  
	\bea
	F^{(r_1,r_2)}_j\equiv \sum_{i\in \text{deI}}\sum_k C^{(r_1,r_2)}_i{\cal B}_{i\to j}^{(k)}s_{11}^k+C^{(r_1,r_2)}_j ,~~~~\forall j\in\text{reI}\ed\label{finite}
	\eea
   to be finite. The $F^{(r_1,r_2)}_j$ is nothing  but the reduction coefficients
   for the degenerate case.  
	Above relation can be written more compactly as 
	\bea C^{(r_1,r_2)}_i\Big\vert_{s_{11}^{-k}}{\cal B}_{i\to j}^{(k)}=- C^{(r_1,r_2)}_j\Big\vert_{s_{11}^{-k}}\label{finite-1}\eea
	where we have treated $C_{j=1,5,6,7}$ and $C_{i=2,3,4}$ as row vector and ${\cal B}$ as a matrix. The symbol $~\Big\vert_{s_{11}^{-k}}$ means to take the coefficients of the $k$-th pole $s_{11}^{-k}$. 
	
	Now we determine decomposition coefficients using the results from rank one to three given in \cite{Feng:2022iuc}.
	\begin{itemize}
		\item For the tensor rank level one, i.e., $r_1+r_2=1$, there are two cases given in \eref{sunsetCr1}. 
		With the only divergent pole $s_{11}^{-1}$, by \eref{finite-1} we get following equations:
		\bea (1,0):~~\left( \begin{array}{ccc} s_{01} & 0 & 0 \end{array}\right)
		\left( \begin{array}{cccc}{\cal B}_{2\to 1}^{(0)} &  {\cal B}_{2\to 5}^{(0)} & {\cal B}_{2\to 6}^{(0)} & {\cal B}_{2\to 7}^{(0)} \\
			{\cal B}_{3\to 1}^{(0)} &  {\cal B}_{3\to 5}^{(0)} & {\cal B}_{3\to 6}^{(0)} & {\cal B}_{3\to 7}^{(0)} \\
			{\cal B}_{4\to 1}^{(0)} &  {\cal B}_{4\to 5}^{(0)} & {\cal B}_{4\to 6}^{(0)} & {\cal B}_{4\to 7}^{(0)} \\
		\end{array}\right)=\left( \begin{array}{cccc} 0 & 0 & 0 & 0 \end{array}\right)\eea
		and 
		\bea (0,1):~~\left( \begin{array}{ccc} 0 & s_{0'1}  & 0 \end{array}\right)
		\left( \begin{array}{cccc}{\cal B}_{2\to 1}^{(0)} &  {\cal B}_{2\to 5}^{(0)} & {\cal B}_{2\to 6}^{(0)} & {\cal B}_{2\to 7}^{(0)} \\
			{\cal B}_{3\to 1}^{(0)} &  {\cal B}_{3\to 5}^{(0)} & {\cal B}_{3\to 6}^{(0)} & {\cal B}_{3\to 7}^{(0)} \\
			{\cal B}_{4\to 1}^{(0)} &  {\cal B}_{4\to 5}^{(0)} & {\cal B}_{4\to 6}^{(0)} & {\cal B}_{4\to 7}^{(0)} \\
		\end{array}\right)=\left( \begin{array}{cccc} 0 & 0 & 0 & 0 \end{array}\right)\eea
		thus we can solve
		\bea {\cal B}^{(0)}_{i\to j} =0\co~~~i=2,3;~~~j=1,5,6,7\ed \label{sun-00}\eea

		\item For the tensor rank level two, i.e., $r_1+r_2=2$, there are three cases $(r_1,r_2)=(2,0),(1,1),(0,2)$. There are also two divergent poles ${1\over s_{11}^2}$ and ${1\over s_{11}}$.  
		Explicitly, we have
		\bea
		\left( \begin{array}{ccc} C^{(2,0)}_2 & C^{(1,1)}_2 & C^{(0,2)}_2\\
			C^{(2,0)}_3 & C^{(1,1)}_3 & C^{(0,2)}_3\\
			C^{(2,0)}_4 & C^{(1,1)}_4  & C^{(0,2)}_4 \end{array}\right)\Bigg\vert_{s_{11}^{-2}}={D\over D-1}\left(
		\begin{array}{ccc}
			M_{12,3} s_{01}^2 & 0 & 2 M_2^2 s_{0'1}^2 \\
			2 M_1^2 s_{01}^2 & 0 & M_{12,3} s_{0'1}^2 \\
			-2 s_{01}^2 & s_{01} s_{0'1} & -2 s_{0'1}^2 \\
		\end{array}
		\right)\co\label{sun-r2-div2}
		\eea
		\bea
		\left( \begin{array}{ccc} C^{(2,0)}_2 & C^{(1,1)}_2 & C^{(0,2)}_2\\
			C^{(2,0)}_3 & C^{(1,1)}_3 & C^{(0,2)}_3\\
			C^{(2,0)}_4 & C^{(1,1)}_4  & C^{(0,2)}_4 \end{array}\right)\Bigg\vert_{s_{11}^{-1}}=\left(
		\begin{array}{ccc}
			\frac{3 D s_{01}^2-M_{12,3} s_{00}}{D-1} & -\frac{2 s_{01} s_{0'1}}{D-1} & \frac{2 \left(D s_{0'1}^2-M_2^2 s_{0'0'}\right)}{D-1} \\
			\frac{2 \left(D s_{01}^2-M_1^2 s_{00}\right)}{D-1} & -\frac{2 s_{01} s_{0'1}}{D-1} & \frac{3 D s_{0'1}^2-M_{12,3} s_{0'0'}}{D-1} \\
			\frac{2 s_{00}}{D-1} & -\frac{s_{0'0}}{D-1} & \frac{2 s_{0'0'}}{D-1} \\
		\end{array}
		\right)\label{sun-r2-div1}
		\eea
		where we have defined
		\bea
		M_{12,3}\equiv M_1^2+M_2^2-M_3^2
		\eea
		for the left hand side of \eref{finite-1}. The corresponding data for the right hand side of \eref{finite-1} is following
		\begin{align}
		&\left( \begin{array}{ccc} C^{(2,0)}_1 & C^{(1,1)}_1 & C^{(0,2)}_1\\
			C^{(2,0)}_5 & C^{(1,1)}_5 & C^{(0,2)}_5\\
			C^{(2,0)}_6 & C^{(1,1)}_6  & C^{(0,2)}_6\\
			C^{(2,0)}_7 & C^{(1,1)}_7  & C^{(0,2)}_7 \end{array}\right)\Bigg\vert_{s_{11}^{-2}},	\left( \begin{array}{ccc} C^{(2,0)}_1 & C^{(1,1)}_1 & C^{(0,2)}_1\\
			C^{(2,0)}_5 & C^{(1,1)}_5 & C^{(0,2)}_5\\
			C^{(2,0)}_6 & C^{(1,1)}_6  & C^{(0,2)}_6\\
			C^{(2,0)}_7 & C^{(1,1)}_7  & C^{(0,2)}_7 \end{array}\right)\Bigg\vert_{s_{11}^{-1}}=	\left(
			\begin{array}{ccc}
				0 & 0 & 0 \\
				0 & 0 & 0 \\
				0 & 0 & 0 \\
				0 & 0 & 0 \\
			\end{array}
			\right),\nn
			&\left(
			\begin{array}{ccc}
				-\frac{2 \left((D+2) M_1^2+D \left(M_2^2-M_3^2\right)\right) s_{01}^2}{D-1} & \frac{2 \left(M_1^2+M_2^2-M_3^2\right) s_{01} s_{0'1}}{D-1} & -\frac{2 \left(D M_1^2+(D+2) M_2^2-D M_3^2\right) s_{0'1}^2}{D-1} \\
				-\frac{4 s_{01}^2}{D-1} & \frac{2 s_{01} s_{0'1}}{D-1} & -\frac{2 D s_{0'1}^2}{D-1} \\
				-\frac{2 D s_{01}^2}{D-1} & \frac{2 s_{01} s_{0'1}}{D-1} & -\frac{4 s_{0'1}^2}{D-1} \\
				\frac{2 D s_{01}^2}{D-1} & -\frac{2 s_{01} s_{0'1}}{D-1} & \frac{2 D s_{0'1}^2}{D-1} \\
			\end{array}
			\right)\ed\label{Cj-divergence}
		\end{align}

		Now we can use the \eref{finite-1} to solve wanted decomposition coefficients.
		
		\begin{itemize}
			\item For the pole $s_{11}^{-2}$
			using the \eref{finite-1} we can solve  
			\begin{align}
				{\cal B}^{(0)}_{i\to j} =0\co~~~i=2,3,4\ed \label{sun-0}
			\end{align}
		It is worth to notice that for $i=2,3$ it is consistent with 
			the result  \eref{sun-00}, but for $i=4$ it is the new solution.

			\item For the pole $s_{11}^{-1}$, using the data \eref{sun-r2-div1} and \eref{Cj-divergence}, by \eref{finite-1} we can solve
			\bea
			{\cal B}_{2\to 1}^{(1)}&=&\frac{2 \left(2 D M_1^2 M_{12,3}-D M_{12,3}^2-6 M_1^2 M_{12,3}+2 M_{12,3}^2+4 M_2^2 M_1^2\right)}{D \left(4 M_1^2 M_2^2-M_{12,3}^2\right)}\co\nn
			{\cal B}_{3\to 1}^{(1)}&=&\frac{2 \left(2 D M_2^2 M_{12,3}-D M_{12,3}^2-6 M_2^2 M_{12,3}+2 M_{12,3}^2+4 M_1^2 M_2^2\right)}{D \left(4 M_1^2 M_2^2-M_{12,3}^2\right)}\co\nn
			{\cal B}_{2\to 5}^{(1)}&=&\frac{4 (D-2) M_1^2}{D \left(4 M_1^2 M_2^2-M_{12,3}^2\right)}\co~
			{\cal B}_{3\to 5}^{(1)}=\frac{2 (D-2) M_{12,3}}{D \left(M_{12,3}^2-4 M_1^2 M_2^2\right)}\co\nn
			{\cal B}_{2\to 6}^{(1)}&=&\frac{2 (D-2) M_{12,3}}{D \left(M_{12,3}^2-4 M_1^2 M_2^2\right)}\co~
			{\cal B}_{3\to 6}^{(1)}=\frac{4 (D-2) M_2^2}{D \left(4 M_1^2 M_2^2-M_{12,3}^2\right)}\co\nn
			{\cal B}_{2\to 7}^{(1)}&=&\frac{2 (D-2) \left(2 M_1^2-M_{12,3}\right)}{D \left(M_{12,3}^2-4 M_1^2 M_2^2\right)}\co
			{\cal B}_{3\to 7}^{(1)}=\frac{2 (D-2) \left(2 M_2^2-M_{12,3}\right)}{D \left(M_{12,3}^2-4 M_1^2 M_2^2\right)}\co\nn 
			{\cal B}_{4\to 1}^{(1)}&=&-\frac{2 M_{12,3}}{D}\co~~
			{\cal B}_{4\to 5}^{(1)}=-\frac{2}{D}\co~~
			{\cal B}_{4\to 6}^{(1)}=-\frac{2}{D}\co~~
			{\cal B}_{4\to 7}^{(1)}=\frac{2}{D}\ed\label{sun-1}
			\eea
			There is one thing we need to emphasize. The expressions of $C$'s have
			different tensor structures of $R$ while the coefficients ${\cal B}$ are
			$R$ independent. Thus the same solution \eref{sun-1} solving  \eref{finite-1} for various tensor structures is very nontrivial thing. 
			Thus for practical purpose, we can focus on  one tensor structure with the simplest expressions to solve ${\cal B}$.

		\end{itemize}

		\item For the tensor rank level 3: $r_1+r_2=3$, there are four cases $(r_1,r_2)=(3,0),(2,1)$, $(1,2),(0,3)$. Now the expressions for $C_{i}, C_j$
		of \eref{finite-1} becomes complicated because of various tensor structures. As we mentioned before, we will give only the part of a particular tensor structure. For example, for $C^{(3,0)}_i$ we give the
		part of the tensor structure $s_{01}^3$ as following
		\bea
		\left( \begin{array}{c} C^{(3,0)}_2 \\ C^{(3,0)}_3 \\ C^{(3,0)}_4 \end{array}\right)\Bigg\vert_{s_{11}^{-3}}=\left(
		\begin{array}{c}
			\frac{(D+2) \left(4 (1-2 D) M_2^2 M_{12,3}+(3 D-2) M_{12,3}^2+8 (D-1) M_1^2 M_2^2\right) s_{01}^3}{(D-1) (3 D-2)} \\
			\frac{2 (D+2) M_1^2 \left((5 D-4) M_{12,3}+4 (1-2 D) M_2^2\right) s_{01}^3}{(D-1) (3 D-2)} \\
			-\frac{2 (D+2) \left((4 D-2) M_{12,3}+(D-2) M_1^2+4 (1-2 D) M_2^2\right) s_{01}^3}{(D-1) (3 D-2)} \\
		\end{array}
		\right)
		\eea 
		while for $C^{(1,2)}_i$ we give the
		part of the tensor structure $s_{01}^2 s_{0'1}$ as following
		\bea
		\left( \begin{array}{c} C^{(1,2)}_2 \\ C^{(1,2)}_3 \\ C^{(1,2)}_4 \end{array}\right)\Bigg\vert_{s_{11}^{-3}}=\left(
		\begin{array}{c}
			\frac{2 (D+2) M_2^2 \left((2 D-1) M_{12,3}-2 (D-1) M_1^2\right) s_{01}^2 s_{0'1}}{(D-1) (3 D-2)} \\
			-\frac{2 (D+2) M_1^2 \left((D-1) M_{12,3}+2 (1-2 D) M_2^2\right) s_{01}^2 s_{0'1}}{(D-1) (3 D-2)} \\
			\frac{(D+2) \left(D M_{12,3}+4 (D-1) M_1^2+4 (1-2 D) M_2^2\right) s_{01}^2 s_{0'1}}{(D-1) (3 D-2)} \\
		\end{array}
		\right)
		\eea
		Expressions of $C_i,C_j$ for other divergent poles and the chosen tensor structures are listed in Appendix \ref{appdix:sunset}.
		
		\begin{itemize}
			\item 
			For the pole $s_{11}^{-3}$, it produce the same  results as given by  \eref{sun-0}.

			\item For the pole $s_{11}^{-2}$, using \eref{finite-1} we find the 
			 same results as given in  \eref{sun-1}.

			\item For the pole $s_{11}^{-1}$, using \eref{finite-1} we find
			 expansion coefficients ${\cal B}_{i\to j}^{(2)}$. Since their expressions are very long, we have collected them into a   attached \textsf{Mathematica} file. To give a flavor, we present ${\cal B}_{4\to 1}^{(2)}$ as following
			 \bea
				& & {\cal B}_{4\to 1}^{(2)}=\frac{1 }{D^2 (D+2) \left(M_{12,3}^2-4 M_1^2 M_2^2\right)^2}\Big[\left(16 D^3-208 D^2+528 D-288\right) M_2^2 M_1^2 M_{12,3}^2\nn
				& &+\big(4 D^3-14 D^2+28 D-48\big) M_{12,3}^4-\big(8 D^3-40 D^2+80 D-
				96\big) (M_1^2+M_2^2) M_{12,3}^3\nn
				&&  +\left(128 D^2-416 D+96\right) (M_2^2 M_1^4+ M_1^2 M_2^4)M_{12,3}+\left(384 D-96 D^2\right) M_2^4 M_1^4\Big]\ed
		    \eea
		\end{itemize}
	\end{itemize}

	~\\
	Now we can calculate the final reduction coefficients by \eref{finite}. For example, the reduction coefficients for rank $(0,1)$ is given by 
	\bea
	F^{(0,1)}_{1}&=&\frac{2 s_{0'1} \left(-2 D M_2^2 M_{12,3}+D M_{12,3}^2+6 M_2^2 M_{12,3}-2 M_{12,3}^2-4 M_1^2 M_2^2\right)}{D \left(M_{12,3}^2-4 M_1^2 M_2^2\right)}\co\nn
	F^{(1,0)}_{1}&=&\frac{2 s_{01} \left(2 D M_1^2 M_{12,3}-D M_{12,3}^2-6 M_1^2 M_{12,3}+2 M_{12,3}^2+4 M_2^2 M_1^2\right)}{D \left(4 M_1^2 M_2^2-M_{12,3}^2\right)}\co\nn
	F^{(0,1)}_{5}&=&\frac{2 (D-2) M_{12,3}s_{0'1}}{D \left(M_{12,3}^2-4 M_1^2 M_2^2\right)}\co\hspace{38pt}
	F^{(1,0)}_{5}=\frac{4 (D-2) M_1^2 s_{01}}{D \left(4 M_1^2 M_2^2-M_{12,3}^2\right)}\co\nn
	F^{(0,1)}_{6}&=&\frac{4 (D-2) M_2^2 s_{0'1}}{D \left(4 M_1^2 M_2^2-M_{12,3}^2\right)}\co\hspace{37pt}
	F^{(1,0)}_{6}=\frac{2 (D-2)M_{12,3}s_{01} }{D \left(M_{12,3}^2-4 M_1^2 M_2^2\right)}\co\nn
	F^{(0,1)}_{7}&=&\frac{2 (D-2) s_{0'1} \left(2 M_2^2-M_{12,3}\right)}{D \left(M_{12,3}^2-4 M_1^2 M_2^2\right)}\co~~
	F^{(1,0)}_{7}=\frac{2 (D-2) s_{01} \left(2 M_1^2-M_{12,3}\right)}{D \left(M_{12,3}^2-4 M_1^2 M_2^2\right)}\ed
	\eea
	In the attached \textsf{Mathematica} file we have present reduction coefficients for other tensor ranks up to rank level two. 
	%
	
	\section{Self-consistence}
	\label{sec:consistence}

	From the previous section we see that  although the decomposition coefficients $\mathcal{B}_{n+1;\what{\mathbf{b}_j}}$ 
	are computed using the tensor reduction results in    \eref{B-m}, 
	by definition in \eref{eq:expand n+1-gon 01}, it should be 
	independent of the auxiliary $R$ and the tensor rank {used in the formula.}  Thus to show the correctness of the formula \eref{B-m}, we should prove this point. 
	
	More explicitly, there are two facts we want to prove. The first one is that to get $\mathcal{B}_{n+1;\what{\mathbf{b}_j}}$ up to $|G|^k$ order, we require $m$ in \eref{B-m} to satisfy condition $m>k$. One can choose different $m_i$ . As long as $m_i>k$, result given by \eref{B-m} should be the same up to $|G|^k$. The second one is that although at the RHS of \eref{B-m}, $C^{(m)}_{n+1\to n+1;\what{\mathbf{b}_j}},\ C^{(m)}_{n+1\to n+1}$ are functions of $R$, as long as $m>k$, the series of $|G|$ should be independent of $R$ up to $|G|^k$.

	In this section, we will {prove these crucial consistencies} using two different methods. 
	\subsection{Proof with recursion relation}
	The {crucial input of our proof is the rather non-trivial relation} for one-loop integrals\footnote{This relation is observed in \cite{Feng:2022iuc}. Here we rephrase it in the language of projective space \cite{Feng:2022rwj}.}
	\bea
	I_{n}^{(m)}&=&{1\over \abs{G}}\left[A_mI_{n}^{(m-1)}+B_mI_{n}^{(m-2)}+\text{Lower Terms}\right]\nn
	&=&{1\over \abs{G}}\left[A_mI_{n}^{(m-1)}+B_mI_{n}^{(m-2)}+\mathcal{O}(1)\right]\ed \label{rankr-deco}
	\eea
	where "$\text{Lower Terms}$" means the contributions from lower topologies, we also denote it as $O(1)$ for it contains no poles of $\abs{G}$, and
	\bea
	A_m&=&a_m(VQ^*L)={2(D+2m-n-2)\over D+m-n-1}(VQ^*L)\co\nn
	B_m&=&b_m\left[(\det Q)R^2-VQ^{*}V)\right]={4(m-1)((\det Q)R^2-VQ^{*}V)\over  D+m-n-1}\ed
	\eea 
	Here $Q^*$ is the adjoint matrix of $Q$ with component $Q_{ij}\equiv {M_i^2+M_j^2-(K_i-K_j)^2\over 2},i,j=0,\ldots, n-1,K_0=0$. The two vectors in the expression of $A_m,B_m$ are $L=\{1,1,\ldots,1\}$, $V=\{0,s_{01},s_{02},\ldots, s_{0,n-1}\}$. One can prove the relation \eref{rankr-deco} by showing that it satisfies the $\cal D$-type and $\cal T$-type differential relations\footnote{Since we have shown the
		reduction coefficients can be completely determined by 
		$\cal D$-type and $\cal T$-type differential relations in \cite{Feng:2021enk,Hu:2021nia}, {checking} relation \eref{rankr-deco} satisfies these two recursions will be sufficient
		for our proof.  }. First, acting with $\cal D$-type operator, we have 
	\bea
	{\cal D}_iI_{n}^{(m)}=m(f_iI_{n}^{(m-1)}+I_{n;\what{0}}^{(m)}-I_{n;\what{i}}^{(m)})=mf_iI_{n}^{(m-1)}+\mathcal{O}(1) \label{Di-deco-LHS}
	\eea
	at one side, while 
	applying ${\cal D}_i$ on the RHS of \eref{rankr-deco}, we find
	\bea
	&&{1\over \abs{G}}{\cal D}_i\left[A_mI_{n}^{(m-1)}+B_mI_{n}^{(m-2)}+\mathcal{O}(1)\right]\nn
	&=&{1\over \abs{G}}\left[({\cal D}_iA_m)I_{n}^{(m-1)}+((m-1)f_i+{\cal D}_iB_m)I_{n}^{(m-2)}+B_m(m-2)f_iI_{n}^{(m-3)}+\mathcal{O}(1)\right]~~~ \label{Di-deco-RHS}
	\eea
	at another side. 
	Identifying  \eref{Di-deco-RHS} and \eref{Di-deco-LHS}, we have
	\bea
	I_{n}^{(m-1)}={1\over mf_i\abs{G}-{\cal D}_iA_m}\left[((m-1)f_i+{\cal D}_iB_m)I_{n}^{(m-2)}+B_m(m-2)f_iI_{n}^{(m-3)}+\mathcal{O}(1)\right]\ed ~~~
	\eea
	By our recursive assumption of the form \eref{rankr-deco},  we just need to prove
	\bea
	&&(mf_i\abs{G}-{\cal D}_iA_m)A_{m-1}=\abs{G}((m-1)f_i+{\cal D}_iB_m),\nn
	&&(mf_i\abs{G}-{\cal D}_iA_m)B_{m-1}=\abs{G}B_m(m-2)f_i\co
	\eea
	which can {be checked using the identities} 
	\bea
	&&{\cal D}_iA_m={2(D+2m-n-2)\over D+m-n-1}\sum_{ab}s_{ia}Q^*_{ab}={2(D+2m-n-2)\over D+m-n-1}f_i\abs{G},\nn
	&& {\cal D}_iB_m={4(m-1)(2(\det Q)s_{0i}-2\sum_{jk}s_{ik}Q^{*}_{kj}s_{0j})\over  D+m-n-1}={4(m-1)f_i(VQ^*L)\over  D+m-n-1}\ed
	\eea
	Next we consider the $\cal T$ operator. First, we have
	\bea
	{\cal T}I_n^{(m)}=4m(m-1)M_0^2I_{n}^{(m-2)}+4m(m-2)I_{n;\what{0}}^{(m-2)}=4m(m-1)M_0^2I_{n}^{(m-2)}+\mathcal{O}(1)\ed
	\eea
	Acting with $\cal T$ on the RHS of \eref{rankr-deco}, we get 
	\bea
	&&{{\cal T}\over |G|}\left[A_mI_{n}^{(m-1)}+B_mI_{n}^{(m-2)}+\mathcal{O}(1)\right]\nn
	&=&{1\over |G|}\left[2a_m(m-1)\sum_{i,j}Q^*_{ij}f_i+2(D+2m-4)b_m\det(Q)-2b_m\sum_{i,j=1}^{n-1}s_{ij}Q^*_{ij}\right]I_n^{(m-2)}\nn
	&&+{1\over |G|}\left[4(m-1)(m-2)M_0^2A_m-4b_m(m-2)\sum_{i,j=1}^{n-1}f_iQ^*_{ij}s_{0j}\right]I_n^{(m-3)}\nn
	&&+{1\over |G|}\left[4b_m(m-2)(m-3)M_0^2(\det(Q)R^2-(VQ^*V))\right]I_n^{(m-4)}
	\eea
	where we have used following algebraic results
	\bea
	&&{\cal T}\left[(VQ^*L)I_n^{(m)}\right]=2\sum_{i,j}Q^*_{ij}mf_iI_{n}^{(m-1)}+4m(m-1)M_0^2(VQ^*L)I_{n}^{(m-2)}+\mathcal{O}(1)\co\nn
	&&{\cal T}\left[R^2I_n^{(m)}\right]=2(D+2m)I_n^{(m)}+4m(m-1)M_0^2R^2I_n^{(m-2)}+\mathcal{O}(1)\co\nn
	&&{\cal T}\left[(VQ^*V)I_n^{(m)}\right]=2\sum_{i,j=1}^{n-1}s_{ij}Q^*_{ij}I_n^{(m)}+4m\sum_{i,j=1}^{n-1}f_iQ^*_{ij}s_{0j}I_{n}^{(m-1)}\nn
	&&\hspace{100pt}+4m(m-1)M_0^2(VQ^*V)I_n^{(m-2)}+\mathcal{O}(1)\ed
	\eea
	Comparing both sides, we just need to prove
	\bea
	&&4m(m-1)M_0^2\abs{G}-\left[2a_m(m-1)\sum_{i,j}Q^*_{ij}f_i+2(D+2m-4)b_m\det(Q)-2b_m\sum_{i,j=1}^{n-1}s_{ij}Q^*_{ij}\right]A_{m-2}\nn
	&&={ -4b_m(m-2)\over |G|}\sum_{i,j=1}^{n-1}f_iQ^*_{ij}s_{0j}
	\eea
	and
	\bea
	&&4m(m-1)M_0^2\abs{G}-\left[2a_m(m-1)\sum_{i,j}Q^*_{ij}f_i+2(D+2m-4)b_m\det(Q)-2b_m\sum_{i,j=1}^{n-1}s_{ij}Q^*_{ij}\right]B_{m-2}\nn
	&=&\left[4a_m(m-1)(m-2)M_0^2(VQ^*L)+4b_m(m-2)(m-3)M_0^2(\det(Q)R^2-(VQ^*V))\right]\ed
	\eea
	One can check the two equations after some algebra.
	
	Now we employ \eref{rankr-deco} to prove self-consistence, i.e., to show that  
	\bea
	\mathcal{B}_{n;\what{\mathbf{b}_j}}= -{\frac{C^{(m)}_{n\to n;\what{\mathbf{b}_j}}}{C^{(m)}_{n\to n}}}+ \mathcal{O}(\abs{G}^m)\label{B-exp}
	\eea
	{is} independent of the rank choice $m$ and $R$.
	First let us show the independence of rank, which is equivalent to the relation
	\bea
	{\frac{C^{(m+1)}_{n\to n;\what{\mathbf{b}_j}}}{C^{(m+1)}_{n\to n}}}-{\frac{C^{(m)}_{n\to n;\what{\mathbf{b}_j}}}{C^{(m)}_{n\to n}}}=\mathcal{O}(\abs{G}^m)\ed
	~~~\label{4-14}\eea
	To see it, first we rewrite \eref{4-14} to 
	\bea
	{C^{(m+1)}_{n\to n;\what{\mathbf{b}_j}}C^{(m)}_{n\to n}-C^{(m)}_{n\to n;\what{\mathbf{b}_j}}C^{(m+1)}_{n\to n}\over C^{(m)}_{n\to n}C^{(m+1)}_{n\to n}}=\mathcal{O}(\abs{G}^m)\ed
	\eea
	Using the divergence behavior of denominator  
	\bea
	C^{(m)}_{n\to n}C^{(m+1)}_{n\to n}=\mathcal{O}(\abs{G}^{-2m-1})\co
	\eea
	we just need to show the leading divergent behavior of numerator is 
	\bea
	{C^{(m+1)}_{n\to n;\what{\mathbf{b}_j}}C^{(m)}_{n\to n}-C^{(m)}_{n\to n;\what{\mathbf{b}_j}}C^{(m+1)}_{n\to n}}=\mathcal{O}(\abs{G}^{-m-1})\ed~~~~\label{4-17}
	\eea
	To prove it, 
	we assume the equation holds for all $m<m'$, while $m'=1,2$ can be checked explicitly, then we prove it holds for $m=m'$. Using \eref{rankr-deco}, the LHS of \eref{4-17} becomes
	\bea
	&&{C^{(m'+1)}_{n\to n;\what{\mathbf{b}_j}}C^{(m')}_{n\to n}-C^{(m')}_{n\to n;\what{\mathbf{b}_j}}C^{(m'+1)}_{n\to n}}\nn
	&=&{1\over \abs{G}}\left[A_{m'+1}C^{(m')}_{n\to n;\what{\mathbf{b}_j}}+B_{m'+1}C^{(m'-1)}_{n\to n;\what{\mathbf{b}_j}}+\mathcal{O}(1)\right]C^{(m')}_{n\to n}\nn
	&&-C^{(m')}_{n\to n;\what{\mathbf{b}_j}}{1\over \abs{G}}\left[A_{m'+1}C^{(m')}_{n\to n}+B_{m'+1}C^{(m'-1)}_{n\to n}+\mathcal{O}(1)\right]\nn
	&=&{B_{m'+1}\over \abs{G}}\left[{C^{(m')}_{n\to n;\what{\mathbf{b}_j}}C^{(m'-1)}_{n\to n}-C^{(m'-1)}_{n\to n;\what{\mathbf{b}_j}}C^{(m')}_{n\to n}}\right]+\mathcal{O}(\abs{G}^{-1})\nn
	&=&\mathcal{O}(\abs{G}^{-m'-1})+\mathcal{O}(\abs{G}^{-1})=\mathcal{O}(\abs{G}^{-m'-1})\ed
	\eea

	Having shown the independence of the rank choice, now 
	we prove the first $m$ terms in \eref{B-exp} don't rely on $R$, i.e.,  the derivation {over $R$ is zero up to order $\mathcal{O}(\abs{G}^m)$:}
	\bea
	{\d\over \d R^{\mu}}\left[{\frac{C^{(m)}_{n\to n;\what{\mathbf{b}_j}}}{C^{(m)}_{n\to n}}}\right]={{\d\over \d R^{\mu}}C^{(m)}_{n\to n;\what{\mathbf{b}_j}}C^{(m)}_{n\to n}-C^{(m)}_{n\to n;\what{\mathbf{b}_j}}{\d\over \d R^{\mu}}C^{(m)}_{n\to n}\over (C^{(m)}_{n\to n})^2}
	\leq \mathcal{O}(\abs{G}^m)\ed
	\eea
	Similarly, since the leading divergent behavior is  $\abs{G}^{-2m}$, we just need to show that 
	\bea
	{{\d\over \d R^{\mu}}C^{(m)}_{n\to n;\what{\mathbf{b}_j}}C^{(m)}_{n\to n}-C^{(m)}_{n\to n;\what{\mathbf{b}_j}}{\d\over \d R^{\mu}}C^{(m)}_{n\to n}}
	\leq \mathcal{O}(\abs{G}^{-m})\ed \label{4-20}
	\eea
	Due to reduction coefficients are functions of $s_{0i},s_{00}$,
	after {acting} ${\d\over \d R^{\mu}}$, the LHS of \eref{4-20} {can be wrote as the form}
	\bea L^\mu\equiv\sum_{i=1}^{n-1} \W\a_i K_i^\mu+\W \a_{n} R^\mu\ed
	~~~\label{L-def}\eea
	{Noticing the limit $|G|\to 0$ means these $K_i$} are not
	linearly independent, thus the limit can be parameterized as
	\bea K_1=\sum_{i=2}^{n-1} a_i K_i+ t K_0 ~~~~\label{K-rel}\eea
	where $t$ is the same order of $|G|$ and $K_0$ is another
	linear independent momentum. Putting it back to \eref{L-def}
	we get
	\bea L^\mu=\sum_{i=2}^{n-1} (\W\a_i+\W \a_1 a_i) K_i^\mu+ \W \a_1  t  K_0^\mu+\W \a_{n} R^\mu\equiv \sum_{i=2}^{n-1} \a_i K_i^\mu+ \a_1  t  K_0^\mu+ \a_{n} R^\mu\ed
	~~~\label{L-def-1}\eea 
	Let us contract $L^\mu$ with $K_0$, $K_i, i=2,...,n-1$
	and $R$, we will get $n$ equations in the matrix form as
	\bea G(K_0,K_2,...,K_{n-1},R) \left( \begin{array}{c} t\a_1  \\ \a_2\\ \vdots \\ \a_{n}\end{array}\right)=\left( \begin{array}{c} b_1  \\ b_2\\ \vdots \\ b_{n}\end{array}\right) \eea
	where $G(K_0,K_2,...,K_{n-1},R)$ is the non-degenerate Gram matrix and 
	\bea 
	b_1 & = & K_0^{\mu}\left[{{\d\over \d R^{\mu}}C^{(m)}_{n\to n;\what{\mathbf{b}_j}}C^{(m)}_{n\to n}-C^{(m)}_{n\to n;\what{\mathbf{b}_j}}{\d\over \d R^{\mu}}C^{(m)}_{n\to n}}\right]\nn 
	b_i &= & K_i^{\mu}\left[{{\d\over \d R^{\mu}}C^{(m)}_{n\to n;\what{\mathbf{b}_j}}C^{(m)}_{n\to n}-C^{(m)}_{n\to n;\what{\mathbf{b}_j}}{\d\over \d R^{\mu}}C^{(m)}_{n\to n}}\right]\nn
	& = & {\cal  D}_iC^{(m)}_{n\to n;\what{\mathbf{b}_j}}C^{(m)}_{n\to n}-C^{(m)}_{n\to n;\what{\mathbf{b}_j}}{\cal  D}_iC^{(m)}_{n\to n},~~~i=2,...,n-1 \nn
	b_n & = & R^\mu\left[{{\d\over \d R^{\mu}}C^{(m)}_{n\to n;\what{\mathbf{b}_j}}C^{(m)}_{n\to n}-C^{(m)}_{n\to n;\what{\mathbf{b}_j}}{\d\over \d R^{\mu}}C^{(m)}_{n\to n}}\right]=0\eea
	where we have used the homogeneity of $R$ for $b_n$.  
	Thus to show $t\a_1$ and all other $\a_i\leq O(\abs{G}^{-m})$, we just need to 
	show $b_i\leq \mathcal{O}(\abs{G}^{-m})$.

	We will {prove} it recursively by  assuming the equation holds for all $m<m'$, where the $m'=1,2$ can be checked explicitly.  For the case $m=m'$, using 
	\bea
	{\cal  D}_iC^{(m)}_{n\to n;\what{\mathbf{b}_j}}=mf_iC^{(m-1)}_{n\to n;\what{\mathbf{b}_j}}+\mathcal{O}(1)\co
	\eea
	we have
	\bea
	b_i&=&{\cal  D}_iC^{(m')}_{n\to n;\what{\mathbf{b}_j}}C^{(m')}_{n\to n}-C^{(m')}_{n\to n;\what{\mathbf{b}_j}}{\cal  D}_iC^{(m')}_{n\to n}\nn
	&=&m'f_i\left[C^{(m'-1)}_{n\to n;\what{\mathbf{b}_j}}C^{(m')}_{n\to n}-C^{(m')}_{n\to n;\what{\mathbf{b}_j}}C^{(m'-1)}_{n\to n}\right]+\mathcal{O}(\abs{G}^{-m'})\nn
	&=&\mathcal{O}(\abs{G}^{-m'})~~~\label{4-27}
	\eea
	where {in the second line} we have used the 
	result \eref{4-17}. {Above argument has shown that $b_i\leq O(\abs{G}^{-m})$ $i=2,..,n-1$ for $m=m'$.} For $b_1$, we need to be more careful. 
	Using \eref{K-rel}, we have 
	\bea b_1 & = & {1\over t}\left( K_1-\sum_{i=2}^{n-1}a_i K_i)\right)\left[{{\d\over \d R^{\mu}}C^{(m)}_{n\to n;\what{\mathbf{b}_j}}C^{(m)}_{n\to n}-C^{(m)}_{n\to n;\what{\mathbf{b}_j}}{\d\over \d R^{\mu}}C^{(m)}_{n\to n}}\right]\ed\eea
	Naively we have ${1\over t}$ in the front, which will make the 
	divergence higher. However, when $t\to 0$, the combination
	$\left( K_1-\sum_{i=2}^{n-1}a_i K_i)\right)\sim t\to 0$, thus the divergent order estimation given in  \eref{4-27} is also
	true for $b_1$. Thus we {finish} the proof for $R$ independence. 
	
	
	\subsection{Another proof}
	
	Now we will give another proof from different point of view. For simplification,  
	we will constrict to the decomposition coefficient $\mathcal{B}_{n+1;\what{\mathbf{b}_j}}$ with $\mathbf{b}_j=\{r+1,\cdots,n\}$. 
	For given tensor rank $m$, the reduction coefficients have the expansion
	\begin{equation}
		\begin{aligned}
			C^{(m)}_{n+1 \to n+1;\widehat{r+1,r+2,...,n}}
			=& \sum_{2a_0+\sum_{k=1}^{n}a_k=m}\Bigg\{ c^{(0,1,\cdots,r)}_{a_1,\cdots,a_{n}}(m) (M_0^2)^{a_0+r-n}\prod_{k=0}^{n} s_{0k}^{a_k}\Bigg\}\\
			C^{(m)}_{n+1 \to n+1}
			=& \sum_{2a_0+\sum_{k=1}^{n}a_k=m}\Bigg\{ c^{(0,1,\cdots,n)}_{a_1,\cdots,a_{n}}(m) (M_0^2)^{a_0+r-n}\prod_{k=0}^{n} s_{0k}^{a_k}\Bigg\}.
		\end{aligned}
	\end{equation}
	Thus we have the expansion of ${\cal F}$ in \eref{gen-dec-2}
	\begin{equation}
		\begin{aligned}
			\mathcal{F}^{(m)}_{n+1\to n+1;\what{r+1,...,n}}=&\sum_{2a_0+\sum a_i=m}\Bigg\{c^{(0,...,r)}_{a_1,...,a_n}(m)+c^{(0,...,n)}_{a_1,...,a_n}(m)\mathcal{B}_{n+1;\what{r+1,...,n}}\Bigg\}(R\cdot R)^{a_0}\prod (R\cdot K_i)^{a_i}\\
			\equiv&\sum_{2a_0+\sum a_i=m}\mathcal{F}^{(m)}_{n+1\to n+1;\what{r+1,...,n}}(a_1,...,a_n)\times(R\cdot R)^{a_0}\prod (R\cdot K_i)^{a_i}.
		\end{aligned}~~~\label{Hu-2}
	\end{equation}
	Since $c^{(0,...,n)}_{a_1,...,a_n}(m)\sim |G|^{-(\frac{m+\sum a_i}{2})}$, smoothness under the limit gives
	%
	\begin{equation}
		\mathcal{B}_{n+1;\what{r+1,...,n}}=-\frac{c^{(0,...,r)}_{a_1,...,a_n}(m)}{c^{(0,...,n)}_{a_1,...,a_n}(m)}+\mathcal{O}(|G|^{\frac{m+\sum a_i}{2}}).~~~\label{Hu-3}
	\end{equation}
	The difference between  \eref{B-m} and \eref{Hu-3} is that we have used expansion coefficients in \eref{Hu-3} instead of 
	the whole  reduction coefficients in \eref{B-m}. Thus  $\mathcal{B}_{n+1;\what{r+1,...,n}}$
	from \eref{Hu-3} will be independent of $R$ manifestly, but we need to show the same result will be obtained taking
	different $a_i'$ and $m$. Overall, the irrelevancy of the tensor rank $m$ and the expansion indices 
	lead to following condition
	\begin{equation}
		\begin{aligned}
			&\frac{c^{(0,...,r)}_{a_1,...,a_n}(m)}{c^{(0,...,n)}_{a_1,...,a_n}(m)}-\frac{c^{(0,...,r)}_{a'_1,...,a'_n}(m')}{c^{(0,...,n)}_{a'_1,...,a'_n}(m')}\leq |G|^{(Min[\frac{m+\sum a_i}{2},\frac{m'+\sum a'_i}{2}])}\\
			\Rightarrow &c^{(0,...,r)}_{a_1,...,a_n}(m)c^{(0,...,n)}_{a'_1,...,a'_n}(m')-c^{(0,...,n)}_{a_1,...,a_n}(m)c^{(0,...,r)}_{a'_1,...,a'_n}(m')\leq |G|^{-Max[\frac{m+\sum a_i}{2},\frac{m'+\sum a'_i}{2}]}
		\end{aligned}\label{small-c}
	\end{equation}
	holds for any nonzero $c^{(0,...,n)}_{a_1,...,a_n}(m),c^{(0,...,n)}_{a'_1,...,a'_n}(m')$.

	Before to prove \eref{small-c}, let us list  several useful relations. Using the ${\cal T}$ operator \cite{Feng:2021enk,Hu:2021nia} we have relation
	\bea
	& &     4m(m+1)c^{(0,1,\cdots,r)}_{a_1,\cdots,a_n}(m-1)
	= (m+1-\sum_{k=1}^n a_k)(D+m+\sum_{k=1}^na_k-1)c^{(0,1,\cdots,r)}_{a_1,\cdots,a_n} (m+1) \nn
	& &+\sum_{0<i<j}2(a_i+1)(a_j+1)\b_{ij}c^{(0,1,\cdots,r)}_{a_1,\cdots,a_i+1,\cdots,a_j+1,\cdots,a_n}(m+1)\nn & &  +\sum_{a=1}^n(a_i+1)(a_i+2)\b_{ii}c^{(0,1,\cdots,r)}_{a_1,\cdots,a_j+2,\cdots,a_n}(m+2).
	\eea
	Using the $\mathcal{D}$-relation \eqref{inverse_formula} twice, we have
	\begin{equation}
		c^{(0,...,r)}_{a_1,...,a_n}(m+1)=\frac{(m+1)\Bigg\{4m\cdot  c^{(0,...,r)}_{a_1,...,a_n}(m-1)-\boldsymbol{\a}^T\widetilde{\boldsymbol{G}}^{-1}{\boldsymbol{O}}^{(0,...,r)}(a_1,...,a_n;m)\Bigg\}}{(m+1-\sum_{k=1}^n a_k)(D+m-n-1)}. 
		\label{Hu-T}
	\end{equation}
	where vector $\boldsymbol{O}^{(0,...,r)}(a_1,...,a_n;m)$ is given in  \eqref{define of O}. Combining with the component form of \eqref{inverse_formula} we have 
	\begin{equation}
		c^{(0,...,r)}_{a_1,...,a_i+1,...,a_n}(m)=\frac{1}{a_i+1}\sum _k \frac{\widetilde{\boldsymbol{G}}^*_{ik}}{|\widetilde{\boldsymbol{G}}|}[\boldsymbol{O}^{(0,...,r)}(a_1,...,a_n;m)]_k.
		~~~\label{Hu-4}
	\end{equation}
	From  \eref{Hu-T} and \eref{Hu-4} we can read
	\begin{equation}
		\begin{aligned}
			&c^{(0,...,n)}_{a_1,...,a_i+1,...,a_n}(m)=\frac{N_i(a_1,...,a_n;m)}{|G|}\left\{A_i\cdot c^{(0,...,n)}_{a_1,...,a_n}(m-1)+\sum_k B_{ik}\cdot c^{(0,...,n)}_{a_1,...,a_k-1,...,a_n}(m)\right\},\\ &c^{(0,...,r)}_{a_1,...,a_i+1,...,a_n}(m)=\frac{N_i(a_1,...,a_n;m)}{|G|}\left\{A_i\cdot c^{(0,...,r)}_{a_1,...,a_n}(m-1)+\sum_k B_{ik}\cdot c^{(0,...,r)}_{a_1,...,a_k-1,...,a_n}(m)\right\}+\frac{X_i^{lower}}{|G|},\\\\
			&c^{(0,...,n)}_{a_1,...,a_n}(m+1)=\frac{M(a_1,...,a_n;m)}{|G|}\left\{D\cdot c^{(0,...,n)}_{a_1,...,a_n}(m-1)+\sum_k E_{k}\cdot c^{(0,...n)}_{a_1,...,a_k-1,...,a_n}(m)\right\},\\
			&c^{(0,...,r)}_{a_1,...,a_n}(m+1)=\frac{M(a_1,...,a_n;m)}{|G|}\left\{D\cdot c^{(0,...,r)}_{a_1,...,a_n}(m-1)+\sum_k E_{k}\cdot c^{(0,...r)}_{a_1,...,a_k-1,...,a_n}(m)\right\}+\frac{Y^{lower}}{|G|}.\\
		\end{aligned}
		\label{relation}
	\end{equation}
	where $A,B_k,D,E_k$ are finite under $|G|\sim |\widetilde{\boldsymbol{G}}|\to 0$. $X_i^{lower},Y^{lower}$ are those contributions from the reduction of lower topologies, which are also finite under $|G|\to 0$. $M_i,N$ are constants decided by $a_1,...,a_n,m$. We also need to mention that under $|G|\to 0$:
	\begin{equation}
		\begin{aligned}
			&A_i:A_j= (\sum_\mu \widetilde{\boldsymbol{G}}^*_{i\mu}\cdot \boldsymbol{\alpha}_\mu):(\sum_\mu \widetilde{\boldsymbol{G}}^*_{j\mu}\cdot \boldsymbol{\alpha}_\mu),\\
			&B_{i{k_i}}:B_{j{k_2}}= \widetilde{\boldsymbol{G}}^*_{ik_1}:\widetilde{\boldsymbol{G}}^*_{ik_2},\\
			&D:E_{k_1}:E_{k_2}= (\sum_{\mu\nu} \boldsymbol{\alpha}_\mu\cdot \widetilde{\boldsymbol{G}}^*_{\mu\nu}\cdot \boldsymbol{\alpha}_\nu):(\sum_\mu \widetilde{\boldsymbol{G}}^*_{k_1\mu}\cdot \boldsymbol{\alpha}_\mu):(\sum_\mu \widetilde{\boldsymbol{G}}^*_{k_2\mu}\cdot \boldsymbol{\alpha}_\mu).
		\end{aligned}
		\label{ABDE}
	\end{equation}
	There are also some useful properties of $\widetilde{\boldsymbol{G}}$. Since $|\widetilde{\boldsymbol{G}}|\to 0$ means the corank of $\widetilde{\boldsymbol{G}}$ is one, we have
	\begin{equation}
		\begin{aligned}
			&\sum_{\mu}\widetilde{\boldsymbol{G}}^*_{i\mu}\cdot \boldsymbol{v_\mu}/\sum_{\mu}\widetilde{\boldsymbol{G}}^*_{j\mu}\cdot \boldsymbol{v_\mu}=\widetilde{\boldsymbol{G}}^*_{ik}/\widetilde{\boldsymbol{G}}^*_{jk},\\
			&\widetilde{\boldsymbol{G}}^*_{ik_1}\widetilde{\boldsymbol{G}}^*_{jk_2}-\widetilde{\boldsymbol{G}}^*_{ik_2}\widetilde{\boldsymbol{G}}^*_{jk_1}=0,\\
			&(\sum_{\mu}\widetilde{\boldsymbol{G}}^*_{i\mu}\cdot \boldsymbol{v_\mu})\cdot (\sum_{\mu}\widetilde{\boldsymbol{G}}^*_{j\mu}\cdot \boldsymbol{v_\mu})=\widetilde{\boldsymbol{G}}_{ij}\cdot \boldsymbol{\a}^T\widetilde{\boldsymbol{G}}^{-1}\boldsymbol{\alpha}.
		\end{aligned}
		\label{popty of G}
	\end{equation}
	
	Having above preparations,  we will prove \eref{small-c} by induction with index $\kappa=Max[\frac{m+\sum a_i}{2},\frac{m'+\sum a'_i}{2}]$. Suppose it is true for $\kappa<\kappa'$, all we need to prove are following four relations:
	\begin{itemize}
		\item (A) With the same $m$  but different $a_i, a_j, a_0$: \textbf{$\frac{c^{(0,...,r)}_{a_1,...,a_i+1,..,a_n}(m)}{c^{(0,...,n)}_{a_1,...,a_i+1,...,a_n}(m)}-\frac{c^{(0,...,r)}_{a_1,..,a_j-1,..,a_n}(m)}{c^{(0,...,n)}_{a_1,...,a_j-1,...,a_n}(m)}\leq |\widetilde{\boldsymbol{G}}|^{(\frac{m+\sum a_i-1}{2})}\\$}\\
		\textbf{Proof of (A):}\\
		Above relation is equal to 
		\begin{equation}
			c^{(0,...,r)}_{a_1,...,a_i+1,..,a_n}(m)c^{(0,...,n)}_{a_1,...,a_j-1,...,a_n}(m)-c^{(0,...,r)}_{a_1,...,a_j-1,..,a_n}(m)c^{(0,...,n)}_{a_1,...,a_i+1,...,a_n}(m)\leq |\widetilde{\boldsymbol{G}}|^{\frac{m+\sum a_i+1}{2}}.
		\end{equation}
		Using \eqref{relation}, we have
		\begin{equation}
			\begin{aligned}
				&c^{(0,...,r)}_{a_1,...,a_i+1,..,a_n}(m)c^{(0,...,n)}_{a_1,...,a_j-1,...,a_n}(m)-c^{(0,...,r)}_{a_1,...,a_j-1,..,a_n}(m)c^{(0,...,n)}_{a_1,...,a_i+1,...,a_n}(m)\\
				=&\frac{N_i(a_1,...,a_n;m)}{|\widetilde{\boldsymbol{G}}|}\times \Bigg\{A_i\cdot[c^{(0,...,r)}_{a_1,...,a_n}(m-1)c^{(0,...,n)}_{a_1,...,a_j-1,...,a_n}(m)-c^{(0,...,n)}_{a_1,...,a_n}(m-1)c^{(0,...,r)}_{a_1,...,a_j-1,...,a_n}(m)]\\
				&+\sum_k B_{ik}\cdot[c^{(0,...,r)}_{a_1,...,a_k-1,...,a_n}(m)c^{(0,...,n)}_{a_1,...,a_j-1,...,a_n}(m)-c^{(0,...,n)}_{a_1,...,a_k-1,...,a_n}(m)c^{(0,...,r)}_{a_1,...,a_j-1,...,a_n}]\Bigg\}\\
				&+\frac{X^{lower}\cdot c^{(0,...,n)}_{a_1,...,a_j-1,...,a_n}(m)}{|\widetilde{\boldsymbol{G}}|} \leq \frac{1}{|\widetilde{\boldsymbol{G}}|}|\widetilde{\boldsymbol{G}}|^{\frac{m+\sum a_i-1}{2}}\sim|\widetilde{\boldsymbol{G}}|^{\frac{m+\sum a_i+1}{2}}.
			\end{aligned} \label{equation of A}
		\end{equation}
		Thus we have done the proof. 
		
		\item (B) With same $a_0$, but different $m, a_i$:   \textbf{$\frac{c^{(0,...,r)}_{a_1,...,a_i+1,..,a_n}(m)}{c^{(0,...,n)}_{a_1,...,a_i+1,...,a_n}(m)}-\frac{c^{(0,...,r)}_{a_1,...,a_n}(m-1)}{c^{(0,...,n)}_{a_1,...,a_n}(m-1)}\leq |\widetilde{\boldsymbol{G}}|^{(\frac{m+\sum a_i}{2})}$}\\
		
		\textbf{Proof of (B):}\\
		Change $c^{(0,...,r)/(0,...,n)}_{a_1,...a_j-1,...,a_m}(m)$ in \eqref{equation of A} into $c^{(0,...,r)/(0,...,n)}_{a_1,...,a_n}(m-1)$, we will reach the proof of (B).
		
		\item (C) With the same $m, a_0$  but different $a_i, a_j$:
		\textbf{$\frac{c^{(0,...,r)}_{a_1,...,a_i+1,..,a_n}(m)}{c^{(0,...,n)}_{a_1,...,a_i+1,...,a_n}(m)}-\frac{c^{(0,...,r)}_{a_1,.,,a_j+1,..,a_n}(m)}{c^{(0,...,n)}_{a_1,...,a_j+1,...,a_n}(m)}\leq |\widetilde{\boldsymbol{G}}|^{(\frac{m+\sum a_i+1}{2})}$}.\\
		
		\textbf{Poof of (C):} 
		
		By \eqref{relation}, we have
		\begin{equation}
			\begin{aligned}
				& c^{(0,...,r)}_{a_1,...,a_i+1,..,a_n}(m)c^{(0,...,n)}_{a_1,...,a_j+1,...,a_n}(m)-c^{(0,...,r)}_{a_1,...,a_j+1,..,a_n}(m)c^{(0,...,n)}_{a_1,...,a_i+1,...,a_n}(m)\\
				=&\frac{N_i(a_1,...,a_n;m)N_j(a_1,...,a_n;m)} {|\widetilde{\boldsymbol{G}}|^2}\times \Bigg\{\sum_k[A_iB_{jk}-A_jB_{ik}]\\
				\times &[c^{(0,...,r)}_{a_1,...,a_n}(m-1)c^{0,...,n)}_{a_1,...,a_k-1,...,a_n}(m)-c^{0,...,r)}_{a_1,...,a_k-1,...,a_n}(m)c^{(0,...,n)}_{a_1,...,a_n}(m-1)]\\
				+&\sum_{k_1<k_2}[B_{ik_1}B_{jk_2}-B_{jk_1}B_{ik_2}]\\
				\times&[c^{(0,...,r)}_{a_1,...,a_{k_1}-1,...,a_n}(m)c^{(0,...,n)}_{a_1,...,a_{k_2}-1,...,a_n}(m)-c^{(0,...,r)}_{a_1,...,a_{k_2-1},...,a_n}(m)c^{(0,...,n)}_{a_1,...,a_{k_1}-1,...,a_n}(m)]\Bigg\}\\
				+&\frac{1}{|\widetilde{\boldsymbol{G}}|}\times\left\{X_i^{lower}C^{(0,...,n)}_{a_1,...,a_j+1,...,a_n}(m)-X_j^{lower}c^{(0,...,n)}_{a_1,...,a_i+1,...,a_n}(m)\right\}.
			\end{aligned}\label{equation C}
		\end{equation}
		From \eqref{ABDE} and \eqref{popty of G}, we know 
		\begin{equation}
			\begin{aligned}
				\lim_{|\widetilde{\boldsymbol{G}}|\to 0}(A_iB_{jk}-A_jB_{ik})=\lim_{|\widetilde{\boldsymbol{G}}|\to 0}(B_{ik_1}B_{jk_2}-B_{jk_1}B_{ik_2})=0.
			\end{aligned}
		\end{equation}
		For the last line of \eqref{equation C}, using the following expression of $X^{lower}_i$ and $c^{(0,...,n)}_{a_1,...,a_i+1,...,a_n}(m)$
		\begin{equation}
			\begin{aligned}
				&c^{(0,...,n)}_{a_1,...,a_i+1,...,a_n}(m)=\frac{1}{(a_i+1)|\widetilde{\boldsymbol{G}}|}\sum_\mu \widetilde{\boldsymbol{G}}^*_{i\mu}\cdot \boldsymbol{O}_\mu^{(0,...,n)}(a_1,...,a_n;m),\\
				&X^{lower}_{i}=-\frac{m }{a_i+1}\sum_\mu \widetilde{\boldsymbol{G}}^*_{i\mu}\cdot \delta_{0a_\mu} c^{(0,...,r)}_{a1,...,\what{a_\mu},...,a_n}(m-1;\what{\mu}),\label{equation of C 01}
			\end{aligned}
		\end{equation}
		then by the first line of \eqref{popty of G}, we know 
		\begin{equation}
			X_i^{lower}C^{(0,...,n)}_{a_1,...,a_j+1,...,a_n}(m)-X_j^{lower}c^{(0,...,n)}_{a_1,...,a_i+1,...,a_n}(m)=0\cdot |\widetilde{\boldsymbol{G}}|^{-\frac{m+\sum a_k+1}{2}}+\mathcal{O}(|\widetilde{\boldsymbol{G}}|^{-\frac{m+\sum a_k-1}{2}}).
		\end{equation}
		Thus we have
		\begin{equation}
			\begin{aligned}
				&c^{(0,...,r)}_{a_1,...,a_i+1,..,a_n}(m)c^{(0,...,n)}_{a_1,...,a_j+1,...,a_n}(m)-c^{(0,...,r)}_{a_1,...,a_j+1,..,a_n}(m)c^{(0,...,n)}_{a_1,...,a_i+1,...,a_n}(m)\\
				\leq& \frac{1}{|\widetilde{\boldsymbol{G}}|^2}\cdot \mathcal{O}(|\widetilde{\boldsymbol{G}}|)\cdot \mathcal{O}(|\widetilde{\boldsymbol{G}}|^{-\frac{m+\sum a_k-1}{2}})
				+\frac{1}{|\widetilde{\boldsymbol{G}}|}\cdot \mathcal{O}(|\widetilde{\boldsymbol{G}}|^{-\frac{m+\sum a_k -1}{2}})\sim |\widetilde{\boldsymbol{G}}|^{-\frac{m+\sum a_k +1}{2}}.
			\end{aligned}
		\end{equation}
		
		\item (D) With the same $a_0$, but different $m, a_i$: \textbf{$\frac{c^{(0,...,r)}_{a_1,...,a_i+1,..,a_n}(m)}{c^{(0,...,n)}_{a_1,...,a_i+1,..,a_n}(m)}-\frac{c^{(0,...,r)}_{a_1,...,a_i,..,a_n}(m+1)}{c^{(0,...,n)}_{a_1,...,a_i,..,a_n}(m+1)}\leq |\widetilde{\boldsymbol{G}}|^{\frac{m+\sum a_i+1}{2}}$}\\
		
		\textbf{Proof of (D):}
		
		By \eqref{relation}, we have
		\begin{equation}
			\begin{aligned}
				&c^{(0,...,r)}_{a_1,...,a_i+1,..,a_n}(m)c^{(0,...,n)}_{a_1,...,a_i,..,a_n}(m+1)-c^{(0,...,r)}_{a_1,...,a_i,..,a_n}(m+1)c^{(0,...,n)}_{a_1,...,a_i+1,..,a_n}(m)\\
				=&\frac{N_i(a_1,...,a_n;m)M(a_1,...,a_n;m)} {|\widetilde{\boldsymbol{G}}|^2}\times \Bigg\{\sum_{k_1<k_2}[E_{k_1}B_{ik_2}-E_{k_2}B_{ik_1}]\\
				\times & [c^{(0,...,r)}_{a_1,...,a_{k_1}-1,...,a_n}(m)c^{(0,...,n)}_{a_1,...,a_{k_2}-1,...,a_n}(m)-c^{(0,...,r)}_{a_1,...,a_{k_2}-1,...,a_n}(m)c^{(0,...,n)}_{a_1,...,a_{k_1}-1,...,a_n}(m)]\\
				+&\sum_k [A_{i}E_{k}-D\cdot B_{ik}]\\
				\times &[c^{(0,...,n)}_{a_1,...,a_n}(m-1)c^{(0,...,r)}_{a_1,...,a_k-1,...,a_n}(m)-c^{(0,...,r)}_{a_1,...,a_n}(m-1)c^{(0,...,n)}_{a_1,...,a_k-1,...,a_n}(m)]\Bigg\}\\
				+&\frac{1}{|\widetilde{\boldsymbol{G}}|}[Y^{lower}c^{(0,...,n)}_{a_1,...,a_i+1,...,a_n}(m)-X^{lower}_ic^{(0,...,n)}_{a_1,...,a_n}(m+1)].
			\end{aligned}\label{equation of D}
		\end{equation}
		Similar to the  proof of (C), we have
		\begin{equation}
			\begin{aligned}
				\lim_{|\widetilde{\boldsymbol{G}}|\to 0}(E_{k_1}B_{ik_2}-E_{k_2}B_{ik_1})=\lim_{|\widetilde{\boldsymbol{G}}|\to 0}(A_{i}E_{k}-D\cdot B_{ik})=0.
			\end{aligned}
		\end{equation}
		Then using the following expressions for $Y^{lower}$ and $c^{(0,...,n)}_{a_1,...,a_n}(m+1)$
		\begin{equation}
			\begin{aligned}
				Y^{lower}=&\frac{m(m+1)}{(m+1-\sum a_\alpha)(D+m-n-1)}\cdot \sum_\mu \widetilde{\boldsymbol{G}}^*_{i\mu}\cdot \delta_{0a_\mu} c^{(0,...,r)}_{a1,...,\what{a_\mu},...,a_n}(m-1;\what{\mu})\\
				c^{(0,...,n)}_{a_1,...,a_n}(m+1)=&\frac{m+1}{(m+1-\sum a_\alpha)(D+m-n-1)|\widetilde{\boldsymbol{G}}|}\times\\ &\Bigg\{4m|\widetilde{\boldsymbol{G}}|\cdot c^{(0,...,n)}_{a_1,...,a_n}(m-1)-\sum_{\mu\nu}\alpha_\nu  \widetilde{\boldsymbol{G}}^*_{\mu\nu}O_\nu^{(0,...,n)}(a_1,...,a_n;m)\Bigg\}
			\end{aligned}
		\end{equation}
		and combining with \eqref{equation of C 01} and the property \eqref{popty of G}, the leading order of last line of \eqref{equation of D} is
		\begin{equation}
			\begin{aligned}
				Y^{lower}c^{(0,...,n)}_{a_1,...,a_i+1,...,a_n}(m)-X^{lower}_ic^{(0,...,n)}_{a_1,...,a_n}(m+1)=0\cdot |\widetilde{\boldsymbol{G}}|^{-\frac{m+\sum a_k+1}{2}}+\mathcal{O}(|\widetilde{\boldsymbol{G}}|^{-\frac{m+\sum a_k-1}{2}}).
			\end{aligned}
		\end{equation}
		thus we arrive 
		\begin{equation}
			\begin{aligned}
				&c^{(0,...,r)}_{a_1,...,a_i+1,..,a_n}(m)c^{(0,...,n)}_{a_1,...,a_i,..,a_n}(m+1)-c^{(0,...,r)}_{a_1,...,a_i,..,a_n}(m+1)c^{(0,...,n)}_{a_1,...,a_i+1,..,a_n}(m)\\
				\leq& \frac{1}{|\widetilde{\boldsymbol{G}}|^2}\cdot \mathcal{O}(|\widetilde{\boldsymbol{G}}|)\cdot \mathcal{O}(|\widetilde{\boldsymbol{G}}|^{-\frac{m+\sum a_k-1}{2}})
				+\frac{1}{|\widetilde{\boldsymbol{G}}|}\cdot \mathcal{O}(|\widetilde{\boldsymbol{G}}|^{-\frac{m+\sum a_k -1}{2}})\sim |\widetilde{\boldsymbol{G}}|^{-\frac{m+\sum a_k +1}{2}}.
			\end{aligned}
		\end{equation}
	\end{itemize}
	Having above four relations (A)(B)(C)(D), we can prove \eqref{small-c} iteratively. Eventually, we have proved that $\mathcal{B}_{n+1;\mathbf{b}_j}$ is independent of the rank $m$ and $R$ by the recursion relation of $c^{(0,...,r)}_{a_1,...,a_n}(m)$. 
	
	\section{Discussion}
	\label{sec:discussion}
	In this paper, we show how to do the reduction for one-loop tensor integrals when Gram determinant degenerates. 
	In this case,  the highest topological master basis will be reduced to an expansion of {the} basis with lower typologies. By demanding the cancellation of divergent parts, we can solve the {decomposition} coefficients. {We have proved that the results are independent of the tensor rank $m$ and the auxiliary vector $R$, thus shown our method's self-consistency.}
	
	One advantage of our method is that we have the series expansion around the degenerate point $|G|$, which will be helpful for some situations, for example, the improvement of numerical accuracy. The same idea can also be used to deal with other kinds of singularities, such as soft limits or massless limits. Furthermore, as the improved PV-reduction method has the applied to two-loop  sunset topology recently in \cite{Feng:2022iuc}, we show that 
	 the same method works well for the sunset topology in the limit $K^2\to 0$ as given in the section \ref{subsec:sunset}.

	\section*{Acknowledgments}
	This work is supported by  Chinese NSF funding under Grant No.11935013, No.11947301, No.12047502 (Peng Huanwu Center).

	\appendix
	%
	
	%

	\section{Expansion for degenerate Gram determinant}
	\label{appdix:A}
	In this appendix, we will give the decomposition for box and
	pentagon as the series of $|G|$.
	Since the analytic expressions with higher and higher orders of $|G|$ are complicated, we will present only numerical results to demonstrate the independence of decomposition coefficients for the rank choice and $R$.

	In following, we list some numerical results for box and pentagon.   Here we set $K_0=0$, and choose the Gram Matrices as: 
	\bea
	\text{box}:
	\mathbf{G}&=&\frac{1}{71}\left(
	\begin{array}{ccc}
		2 & 5  & 11 \\
		5 & 17 & 23  \\
		11  & 23 & 65+\frac{357911}{9}t  \\
	\end{array}
	\right)\co\nn
	\text{pentagon}:
	\mathbf{G}&=&\frac{1}{71}\left(
	\begin{array}{cccc}
		2 & 5  & 11 & 41 \\
		5 & 17 & 23 & 47 \\
		11 & 23 & 31 & 59 \\
		41  & 47 & 59 &\frac{1283}{17}+\frac{25411681}{306}t  \\
	\end{array}
	\right)\co
	\eea
	where $\mathbf{G}_{ij}=s_{ij}=K_i\cdot K_j$. With  this parameterization, we have $\text{det}(\mathbf{G})=t$.
	The chosen squares of masses are
	\bea
	\text{box}:
	\left(\begin{array}{cccc}
		M_0^{2},&M_1^{2},&M_2^{2},&M_3^{2}
	\end{array}\right)&=&\frac{1}{71}
	\left(\begin{array}{cccc}
		3, & 7, & 13, & 19 
	\end{array}\right)\co\nn
	\text{pentagon}:
	\left(\begin{array}{ccccc}
		M_0^{2},&M_1^{2},&M_2^{2},&M_3^{2},&M_4^{2}
	\end{array}\right)&=&\frac{1}{71}
	\left(\begin{array}{ccccc}
		3, & 7,  & 13,& 19, & 29
	\end{array}\right)\ed
	\eea
	
	Firstly we present some reduction coefficients needed in \eref{B-m} and \eref{F-m-2}. For $I_{4}^{(2)}$ we have 
	\begin{subequations}
		\allowdisplaybreaks
		\bea
		\ctoself{2}{4}&=&\frac{ s_{03} s_{01}}{384300851763 (D-3) t^2}\times\big(-12041426688574 D t^2-32933538576 D t-\nn
		&&20155392 D+29975466437514 t^2+73134102096 t+40310784\big)\nn
		&&+\frac{ s_{03}^2}{128100283921 (D-3) t^2}\times\big(128100283921 D t^2+463852656 D t+\nn
		&&419904 D-256200567842 t^2-856838934 t-839808\big)\nn
		&&+\frac{ s_{02} s_{03}}{384300851763 (D-3) t^2}\times\big(2818206246262 D t^2+6493937184 D t+\nn
		&&2519424 D-7686017035260 t^2-16273497348 t-5038848\big)\nn
		&&+\frac{ s_{01}^2}{10376122997601 (D-3) t^2}\times\big(2546761744633401 D t^2+\nn
		&&4709032163712 D t+2176782336 D+779424512213633527 t^3-\nn
		&&3722722351028181 t^2-8800212589632 t-4353564672\big)\nn
		&&+\frac{ s_{02} s_{01}}{10376122997601 (D-3) t^2}\times\big(-1192101242168826 D t^2-\nn
		&&1690742931120 D t-544195584 D-458485007184490310 t^3+\nn
		&&1496467516765122 t^2+2876350319856 t+1088391168\big)\nn
		&&+\frac{ s_{02}^2}{10376122997601 (D-3) t^2}\times\big(139501209189969 D t^2+\nn
		&&137764238832 D t+34012224 D+91697001436898062 t^3-\nn
		&&25363856216358 t^2-69403953654 t-68024448\big)\nn
		&&+\frac{s_{00}\left(-128100283921 t^2-534719034 t-419904\right) }{228705129 (D-3) t}\co
		\eea
		
		\bea
		\ctonext{2}{4}{3}&=&\frac{ s_{02} s_{01}}{16238064159 (D-3) t^2}\times\big(2145318534 D t+839808 D+\nn
		&&1281002839210 t^2-3652839666 t-1679616\big)+\frac{(357911 t+648) s_{00}}{3t57911 (D-3) }\nn
		&&+\frac{s_{03} s_{01}}{1804229351 (D-3) t^2}\times(100930902 D t+93312 D-251253522 t-\nn
		&&186624)+\frac{ s_{02}^2}{16238064159 (D-3) t^2}\times\big(-183608343 D t-52488 D-\nn
		&&256200567842 t^2+57981582 t+104976\big)+\frac{ s_{01}^2}{16238064159 (D-3) t^2}\times\nn
		&&\big(-5411614320 D t-3359232 D-2177704826657 t^2+10436684760 t+\nn
		&&6718464\big)+\frac{ s_{03}^2(-3221199 D t-5832 D+6442398 t+11664)}{1804229351 (D-3) t^2}\nn
		&&+\frac{ s_{02} s_{03}(-23622126 D t-11664 D+64423980 t+23328)}{1804229351 (D-3) t^2}\co
		\eea
		\bea
		\ctonext{2}{4}{23}&=&\frac{s_{01}^2(48317985 t+10368) }{71 t (715822 t+81)}-\frac{32 s_{02} s_{01}}{71 t}+\frac{s_{03} s_{01}(-7158220 t-2592) }{71 t (715822 t+81)}+\frac{2 s_{02}^2}{71 t}\nn
		&&+\frac{162 s_{03}^2}{71 t (715822 t+81)}+\frac{4 s_{02} s_{03}}{71 t}\ed
		\eea
	\end{subequations}
	For $I_{5}^{(2)}$, we have 
	\begin{subequations}
		\allowdisplaybreaks
		\begin{align}
			\ctoself{2}{5}=&\frac{ s_{04} s_{01}}{32933430093533811 (D-4) t^2}\times\big(143357283936558942 D t^2+23554798645968 D t+\nn
			&801715968 D-514019810871625756 t^2-65771935818336 t-2405147904\big)\nn
			&+\frac{ s_{04}^2}{645753531245761 (D-4) t^2}\times\big(645753531245761 D t^2+150030564624 D t+\nn
			&8714304 D-1937260593737283 t^2-290912924088 t-26142912\big)\nn
			&+\frac{ s_{00}}{552094181406 (D-4) t}\times
			\left(645753531245761 t^2+309209334408 t
			+8714304\right)\nn
			&+\frac{ s_{02} s_{04}}{32933430093533811 (D-4) t^2}\times\big(484315148434320750 D t^2+100670508862704 D t+\nn
			&5158867968 D-1554974503239792488 t^2-219630109481280 t-15476603904\big)\nn
			&-\frac{ s_{03} s_{04}}{10977810031177937 (D-4) t^2}\times\big(167895918123897860 D t^2+36157366074384 D t+\nn
			&1934575488 D-528226388559032498 t^2-75984992058960 t-5803726464\big)\nn
			&+\frac{ s_{01}^2}{30232888825864038498 (D-4) t^2}\times\big(143213926652622383058 D t^2+\nn
			&13789009133462592 D t+331910410752 D-16409682740640811134241 t^3-\nn
			&556117775080599407112 t^2-43600316518303488 t-995731232256\big)\nn
			&+\frac{ s_{02} s_{01}}{15116444412932019249 (D-4) t^2}\times\big(483830833285886429250 D t^2+\nn
			&67656883239451296 D t+2135771338752 D+804074454291399745577809 t^3-\nn
			&1392037972235862072480 t^2-183993225050551104 t-6407314016256\big)\nn
			&+\frac{ s_{03} s_{01}}{1679604934770224361 (D-4) t^2}\times\big(-55909340735257987380 D t^2-\nn
			&8237128089551472 D t-266971417344 D-65638730962563244536964 t^3+\nn
			&172423941884993136132 t^2+22589794735742304 t+800914252032\big)\nn
			&+\frac{ s_{02}^2}{60465777651728076996 (D-4) t^2}\times\big(3269127251931665062500 D t^2+\nn
			&599522136237504000 D t+27486448533504 D-968171281697807856920219 t^3-\nn
			&11503282926679737573096 t^2-1435853145074044608 t-82459345600512\big)\nn
			&+\frac{s_{03}^2}{6718419739080897444 (D-4) t^2}\times\big(392876448409920992400 D t^2+\nn
			&77937877710875520 D t+3865281825024 D+16409682740640811134241 t^3-\nn
			&1277220411366240783636 t^2-174378410197442208 t-11595845475072\big)\nn
			& +\frac{ s_{02} s_{03}}{3359209869540448722 (D-4) t^2}\times\big(-377765815778770185000 D t^2-\nn
			&72109190275233120 D t-3435806066688 D+16409682740640811134241 t^3+\nn
			&1264318255811950478856 t^2+166745623518606528 t+10307418200064\big)\co		
		\end{align}
		\bea
		\ctonext{2}{5}{4}&=&\frac{ s_{01}^2}{1391553384233823 (D-4) t^2}\times\big(3703193448768 D t+112435776 D-\nn
		&&645753531245761 t^2-13919908971456 t-337307328\big)\nn
		&&+\frac{ s_{02} s_{01}}{1391553384233823 (D-4) t^2}\times\big(33381800628840 D t+1446999552 D+\nn
		&&63283846062084578 t^2-113719508702928 t-4340998656\big)\nn
		&&+\frac{ s_{03} s_{01}}{154617042692647 (D-4) t^2}\times\big(-4023685569540 D t-180874944 D-\nn
		&&5166028249966088 t^2+13937595501432 t+542624832\big)\nn
		&&+\frac{s_{04} s_{01}(33848359092 D t+1629504 D-121366188456 t-4888512)}{9095120158391 (D-4) t^2}\nn
		&&+\frac{ s_{02}^2}{2783106768467646 (D-4) t^2}\times\big(122937240222144 D t+9311127552 D-\nn
		&&38099458343499899 t^2-416021948215128 t-27933382656\big)\nn
		&&+\frac{ s_{03}^2}{309234085385294 (D-4) t^2}\times\big(15130216514124 D t+1309377312 D+\nn
		&&645753531245761 t^2-49174347196548 t-3928131936\big)\nn
		&&+\frac{s_{04}^2(7775974386 D t+903312 D-23327923158 t-2709936) }{9095120158391 (D-4) t^2}\nn
		&&+\frac{ s_{02} s_{03}}{154617042692647 (D-4) t^2}\times\big(-14408118186828 D t-1163890944 D+\nn
		&&645753531245761 t^2+47688373738392 t+3491672832\big)\nn
		&&+\frac{s_{02} s_{04}(114352564500 D t+10485504 D-367147967088 t-31456512) }{9095120158391 (D-4) t^2}\nn
		&&+\frac{ s_{03} s_{04}(-118926667080 D t-11796192 D+374161591044 t+35388576)}{9095120158391 (D-4) t^2}\nn
		&&+\frac{s_{00}(25411681 t+2952)}{25411681 (D-4) t}\co
		\eea
		\bea
		\ctonext{2}{5}{34}&=&\frac{s_{01}^2(1568612244768 t-8455536) }{85697 t (25411681 t+443556)}+\frac{s_{02} s_{01}(-760215848796 t-108819072) }{85697 t (25411681 t+443556)}\nn
		&&+\frac{276 s_{03} s_{01}}{85697 t}-\frac{999 s_{03}^2}{85697 t}+\frac{1776 s_{02} s_{03}}{85697 t}+\frac{s_{04} s_{01}(-3659282064 t-1102896) }{5041 t (25411681 t+443556)}\nn
		&&+\frac{s_{02}^2(70517414775 t-350113536) }{85697 t (25411681 t+443556)}-\frac{611388 s_{04}^2}{5041 t (25411681 t+443556)}\nn
		&&+\frac{s_{02} s_{04}(457410258 t-7096896) }{5041 t (25411681 t+443556)}+\frac{18 s_{03} s_{04}}{5041 t}\ed
		\eea
	\end{subequations}

	When $t=\text{det}(\mathbf{G})\to0$, the basis with the highest typology is decomposed to the combinations of lower ones. 
	Let us consider the coefficients  $\mathcal{B}_{4;\what{3}}$ and $\mathcal{B}_{5;\what{4}}$ by \eqref{B-m} for different $m$:
	
	\begin{itemize}
		\item \textbf{$\mathcal{B}_{4;\what{3}}$}
		\begin{itemize}
			
			\item {$m=2$:}

			\bea 
			\mathcal{B}_{4;\what{3}}&=&-{C^{(2)}_{4 \to 4;\what{3}}\over C^{(2)}_{4\to 4}}+\mathcal{O}\left(t^2\right)= \frac{71}{72}-\frac{25411681 (36 D-61) t}{1679616 (D-2)}+\mathcal{O}\left(t^2\right)\ed
			\eea

			\item {$m=3$:}

			\bea \mathcal{B}_{4;\what{3}}&=&-{C^{(3)}_{4 \to 4;\what{3}}\over C^{(3)}_{4\to 4}}+\mathcal{O}\left(t^3\right)= \frac{71}{72}-\frac{25411681 (36 D-61) t}{1679616 (D-2)}\nn & & +\frac{9095120158391 \left(432 D^2-468 D+121\right) t^2}{13060694016 ((D-2) D)}+\mathcal{O}\left(t^3\right)\ed
			\eea
			\item {$m=4$:} 
			\bea \mathcal{B}_{4;\what{3}}&=&-{C^{(4)}_{4 \to 4;\what{3}}\over C^{(4)}_{4\to 4}}+\mathcal{O}\left(t^4\right)= \frac{71}{72}-\frac{25411681 (36 D-61) t}{1679616 (D-2)}\nn & & +\frac{9095120158391 \left(432 D^2-468 D+121\right) t^2}{13060694016 ((D-2) D)}\nn
			& & - \frac{3255243551009881201 \left(15552 D^3+28512 D^2+16596 D+50215\right) t^3}{304679870005248 D \left(D^2-4\right)}\nn
			&&+\mathcal{O}\left(t^4\right)\ed
			\eea
			
			
		\end{itemize}
		\item \textbf{ $\mathcal{B}_{5;\what{4}}$}
		\begin{itemize}
			
			\item {$m=2$:}

			\bea
			\mathcal{B}_{5;\what{4}}&=&-{C^{(2)}_{5 \to 5;\what{4}}\over C^{(2)}_{5\to 5}}+\mathcal{O}\left(t^2\right)=\frac{1207}{164}-\frac{30671898967 (41 D-36) t}{19849248 (D-3)}+\mathcal{O}\left(t^2\right)\ed\nn
			\eea

			\item {$m=3$:}
			\bea
			\mathcal{B}_{5;\what{4}}&=&-{C^{(3)}_{5 \to 5;\what{4}}\over C^{(3)}_{5\to 5}}+\mathcal{O}\left(t^3\right)=\frac{1207}{164}-\frac{30671898967 (41 D-36) t}{19849248 (D-3)}\nn
			&&+\frac{779424512213633527 \left(1681 D^2+3977 D+17049\right) t^2}{2402394183936 (D-3) (D-1)}+\mathcal{O}\left(t^3\right)\ed\nn
			\eea
			
			\item {$m=4$:} 
			\bea
			\mathcal{B}_{5;\what{4}}&=&-{C^{(4)}_{5 \to 5;\what{4}}\over C^{(4)}_{5\to 5}}+\mathcal{O}\left(t^4\right)=\frac{1207}{164}-\frac{30671898967 (41 D-36) t}{19849248 (D-3)}\nn
			&&+\frac{779424512213633527 \left(1681 D^2+3977 D+17049\right) t^2}{2402394183936 (D-3) (D-1)}\nn
			&&-\frac{19806487067953459039028887 t^3}{290766572870141952 ((D-3) (D-1) (D+1))}\times\nn
			&&\big(68921 D^3+670719 D^2+4586014 D+13861761\big)+\mathcal{O}\left(t^4\right)\ed
			\eea

		\end{itemize}
	\end{itemize}
	To make them look more intuitive, we choose $D=4$, and here we list:
	
	\begin{table}[H]
		\centering
		\caption{$\mathcal{B}_{4;\what{3}}$}
		\renewcommand{\arraystretch}{1.5}
		\begin{tabular}{|>{\centering\arraybackslash}p{2.15cm}|>{\centering\arraybackslash}p{1.0cm}|>{\centering\arraybackslash}p{2.0cm}|>{\centering\arraybackslash}p{2.8cm}|>{\centering\arraybackslash}p{4.0cm}|}
			\hline
			\diagbox{rank}{order} & $t^0$ & $t^1$ & $t^2$ &$t^3$\\
			\hline
			$m=2$ & $-\frac{71}{72}$ & $\frac{2109169523 }{3359232}$ & \slash &\slash  \\
			\hline
			$m=3$ & $-\frac{71}{72}$ & $\frac{2109169523 }{3359232}$ & $-\frac{46939915137455951}{104485552128}$ & \slash \\ 
			\hline
			$m=4$ & $-\frac{71}{72}$ & $\frac{2109169523 }{3359232}$ & $-\frac{46939915137455951}{104485552128}$ & $\frac{5104609261966063899030919}{14624633760251904}$ \\ 
			\hline
		\end{tabular}

		\label{number-a-Teqs}
	\end{table}
	\begin{table}[H]
		\centering
		\caption{$\mathcal{B}_{5;\what{4}}$}
		\renewcommand{\arraystretch}{1.5}
		\begin{tabular}{|>{\centering\arraybackslash}p{2.15cm}|>{\centering\arraybackslash}p{1.0cm}|>{\centering\arraybackslash}p{2.0cm}|>{\centering\arraybackslash}p{3.4cm}|>{\centering\arraybackslash}p{5.0cm}|}
			\hline
			\diagbox{rank}{order} & $t^0$ & $t^1$ & $t^2$ &$t^3$\\
			\hline
			$m=2$ & $\frac{1207}{164}$ & $-\frac{122687595868}{620289}$ & \slash &\slash  \\
			\hline
			$m=3$ & $\frac{1207}{164}$ & $-\frac{122687595868}{620289}$ & $\frac{15550298443174202497177}{2402394183936}$ & \slash \\ 
			\hline
			$m=4$ & $\frac{1207}{164}$ & $-\frac{122687595868}{620289}$ & $\frac{15550298443174202497177}{2402394183936}$ & $-\frac{62520186560835559083105672288737}{290766572870141952}$ \\ 
			\hline
		\end{tabular}

		\label{number-a-Teqs}
	\end{table}

	{From these numerical results, one can see that as the series of $t$, $\mathcal{B}_{4;\what{3}}$ and $\mathcal{B}_{5;\what{4}}$ are independent of the choice of rank $m$. And they are also independent of the auxiliary vector $R$, although $s_{0i}=R\cdot K_i$ and $s_{00}=R\cdot R$ appear in the expression of $C^{(3)}_{4 \to 4;\what{3}}$, $C^{(3)}_{4 \to 4}$, etc.}
	
	We can also use $\eqref{F-m-2}$ to derive $F^{(m)}_{n+1;\what{\mathbf{b}_j}}$.  Some results are:
	\begin{subequations}
		\allowdisplaybreaks
		\bea
		F^{(3)}_{4;\what{3}}&=&\frac{\left(-1239447 D^2-1774432 D+652916\right) s_{03} s_{01}^2}{23328 D \left(D^2-4\right)}+\frac{(-197593 D-206184) s_{03}^2 s_{01}}{46656 D \left(D^2-4\right)}\nn
		&&+\frac{\left(862224 D^2+1492491 D-274912\right) s_{02} s_{03} s_{01}}{23328 D \left(D^2-4\right)}-\frac{94501 s_{03}^3}{559872 D \left(D^2-4\right)}\nn
		&&+\frac{(274912 D+455323) s_{02} s_{03}^2}{186624 D \left(D^2-4\right)}+\frac{\left(-1199616 D^2-2674144 D-644325\right) s_{02}^2 s_{03}}{186624 D \left(D^2-4\right)}\nn
		&&+\frac{\left(-7774713 D^3-9053352 D^2+13507608 D+2542936\right) s_{01}^3}{17496 D \left(D^2-4\right)}\nn
		&&+\frac{\left(10816992 D^3+12443673 D^2-19600544 D-3195852\right) s_{02} s_{01}^2}{23328 D \left(D^2-4\right)}\nn
		&&+\frac{\left(-7524864 D^3-10046784 D^2+11389039 D+2955304\right) s_{02}^2 s_{01}}{46656 D \left(D^2-4\right)}\nn
		&&+\frac{\left(10469376 D^3+17994240 D^2-7538496 D-3393445\right) s_{02}^3}{559872 D \left(D^2-4\right)}\nn
		&&+\frac{(759 D-242) s_{00} s_{01}}{108 (D-2) D}+\frac{(121-1056 D) s_{00} s_{02}}{432 (D-2) D}+\frac{121 s_{00} s_{03}}{432 (D-2) D}\co
		\eea
		\bea
		F^{(3)}_{4;\what{23}}&=&\frac{\left(-222655929 D^2+225327659 D+51781152\right) s_{02} s_{03} s_{01}}{2916 D \left(D^2-4\right)}\nn
		&&+\frac{\left(159295600 D^2-46664537 D-85762533\right) s_{02} s_{03}^2}{23328 D \left(D^2-4\right)}+\frac{(28045 D-22791) s_{00} s_{03}}{54 (D-2) D}\nn
		&&+\frac{\left(887216 D^2-259919001 D+121362075\right) s_{02}^2 s_{03}}{23328 D \left(D^2-4\right)}+\frac{(781 D-22791) s_{00} s_{02}}{54 (D-2) D}\nn
		&&+\frac{\left(1053906747 D^3-444555708 D^2-186234704 D-478975656\right) s_{01}^3}{2187 D \left(D^2-4\right)}\nn
		&&+\frac{\left(-8000067 D^3+607223737 D^2-1031570076 D+601955892\right) s_{02} s_{01}^2}{2916 D \left(D^2-4\right)}\nn
		&&+\frac{\left(-752051667 D^3+144661577 D^2+657043940 D-122980236\right) s_{03} s_{01}^2}{2916 D \left(D^2-4\right)}\nn
		&&+\frac{\left(5565264 D^3+9169579 D^2+362503351 D-556647384\right) s_{02}^2 s_{01}}{5832 D \left(D^2-4\right)}\nn
		&&+\frac{\left(269794320 D^3+32842115 D^2-349326177 D+38835864\right) s_{03}^2 s_{01}}{5832 D \left(D^2-4\right)}\nn
		&&+\frac{\left(-7742976 D^3-15727920 D^2-3271609 D+639173595\right) s_{02}^3}{69984 D \left(D^2-4\right)}\nn
		&&+\frac{\left(-193574400 D^3-97997040 D^2+264828935 D+17799771\right) s_{03}^3}{69984 D \left(D^2-4\right)}\nn
		&&+\frac{(91164-79378 D) s_{00} s_{01}}{27 (D-2) D}\co
		\eea
		\bea
		F^{(3)}_{4;\what{123}}&=&\frac{\left(117036897 D^2+80887886 D+52970828\right) s_{01}^3}{486 D (D+2)}-\frac{5041 s_{00} s_{01}}{3 D}\nn
		&&+\frac{\left(-8231953 D^2+40086032 D-66571446\right) s_{02} s_{01}^2}{648 D (D+2)}+\frac{5041 s_{00} s_{02}}{24 D}\nn
		&&+\frac{\left(-79814153 D^2-62992336 D+13600618\right) s_{03} s_{01}^2}{648 D (D+2)}+\frac{5041 s_{00} s_{03}}{24 D}\nn
		&&+\frac{\left(22906304 D^2+79098331 D+246242768\right) s_{02}^2 s_{01}}{5184 D (D+2)}\nn
		&&+\frac{\left(114531520 D^2+130637515 D-17179728\right) s_{03}^2 s_{01}}{5184 D (D+2)}\nn
		&&+\frac{\left(-1206875892 D^2-5693290277 D-9613489460\right) s_{02}^3}{2115072 D (D+2)}\nn
		&&+\frac{\left(-10269182412 D^2-15210859589 D-1023625460\right) s_{03}^3}{8087040 D (D+2)}\nn
		&&+\frac{\left(-10021508 D^2+107731211 D+75877132\right) s_{02} s_{03}^2}{41472 D (D+2)}\nn
		&&+\frac{\left(10021508 D^2+56192027 D-107373300\right) s_{02}^2 s_{03}}{41472 D (D+2)}\nn
		&&+\frac{(-101288813 D-22906304) s_{02} s_{03} s_{01}}{2592 D (D+2)}\co
		\eea
		\bea
		F^{(3)}_{5;\what{4}}&=&\frac{ s_{01}^3}{88197652332 (D-3) \left(D^2-1\right)}\times\big(-1343847503375 D^3+10417729271235 D^2-\nn
		&&17337620658673 D-1794280981875\big)+\frac{ s_{02} s_{01}^2}{29399217444 (D-3) \left(D^2-1\right)}\times\nn
		&&\big(-1633291581025 D^3+12077972337027 D^2-68514754134455 D+\nn
		&&93475183460901\big)+\frac{ s_{03} s_{01}^2}{9799739148 (D-3) \left(D^2-1\right)}\times\big(392816962525 D^3\nn
		&&-4603678621302 D^2+27041270862119 D-33850184121582\big)\nn
		&&+\frac{\left(14623481275 D^2-89044148172 D+94194083969\right) s_{04} s_{01}^2}{192151748 (D-3) \left(D^2-1\right)}\nn
		&&+\frac{ s_{02}^2 s_{01}}{29399217444 (D-3) \left(D^2-1\right)}\times\big(-1985077460015 D^3+30582780834201 D^2-\nn
		&&243862525327933 D+854168652203259\big)+\frac{(580029-231855 D) s_{00} s_{01}}{28577 (D-3) (D-1)}\nn
		&&+\frac{(5019844059-1432168335 D) s_{04}^2 s_{01}}{5651522 (D-3) \left(D^2-1\right)}+\frac{ s_{03}^2 s_{01}}{3266579716 (D-3) \left(D^2-1\right)}\times\nn
		&&\big(-114823419815 D^3+2531886930939 D^2-34017449607547 D+\nn
		&&122326329454647\big)+\frac{ s_{02} s_{03} s_{01}}{4899869574 (D-3) \left(D^2-1\right)}\times\big(477423692915 D^3-\nn
		&&8829762425970 D^2+90569439300937 D-324098013014706\big)\nn
		&&+\frac{\left(17773154165 D^2-276974011536 D+978449946715\right) s_{02} s_{04} s_{01}}{96075874 (D-3) \left(D^2-1\right)}\nn
		&&+\frac{\left(-12823668195 D^2+313094031390 D-1106429051883\right) s_{03} s_{04} s_{01}}{96075874 (D-3) \left(D^2-1\right)}\nn
		&&+\frac{ s_{02}^3}{88197652332 (D-3) \left(D^2-1\right)}\times\big(-2412632605249 D^3+76689368763465 D^2-\nn
		&&1007290897198703 D+6188766114569991\big)+\frac{ s_{03}^3}{3266579716 (D-3) \left(D^2-1\right)}\times\nn
		&&\big(33563768869 D^3-1300394788050 D^2+30312717181523 D-352004287844976\big)\nn
		&&+\frac{2384439363 s_{04}^3}{5651522 (D-3) \left(D^2-1\right)}+\frac{ s_{02} s_{03}^2}{3266579716 (D-3) \left(D^2-1\right)}\times\nn
		&&\big(-139554617929 D^3+5018030800863 D^2-98440205350997 D+\nn
		&&919471445452659\big)+\frac{(39776661783-1740635361 D) s_{02} s_{04}^2}{5651522 (D-3) \left(D^2-1\right)}\nn
		&&+\frac{(1255901463 D-45451057824) s_{03} s_{04}^2}{5651522 (D-3) \left(D^2-1\right)}+\frac{(2522217-281793 D) s_{00} s_{02}}{28577 (D-3) (D-1)}\nn
		&&+\frac{ s_{02}^2 s_{03}}{9799739148 (D-3) \left(D^2-1\right)}\times\big(580253411389 D^3-19518756588588 D^2+\nn
		&&316580122515497 D-2391505497411054\big)+\frac{(203319 D-2723796) s_{00} s_{03}}{28577 (D-3) (D-1)}\nn
		&&+\frac{\left(21601218139 D^2-722503219422 D+7448642198855\right) s_{02}^2 s_{04}}{192151748 (D-3) \left(D^2-1\right)}\nn
		&&+\frac{\left(11245370571 D^2-552200043258 D+9805499547885\right) s_{03}^2 s_{04}}{192151748 (D-3) \left(D^2-1\right)}\nn
		&&+\frac{\left(-15585689037 D^2+639672478488 D-8561581798443\right) s_{02} s_{03} s_{04}}{96075874 (D-3) \left(D^2-1\right)}\nn
		&&+\frac{22707 s_{00} s_{04}}{1681 (D-3) (D-1)}\co
		\eea
		\begin{align}
			F^{(3)}_{5;\what{34}}=&\frac{ s_{01}^3}{36248334956758812852 (D-3) \left(D^2-1\right)}\times\big(-8299431779977556629537 D^3+\nn
			&22726783007443171457382 D^2+12750306686537356621621 D-\nn
			&25457369478990513141690\big)+\frac{ s_{02} s_{01}^2}{81640394046754083 (D-3) \left(D^2-1\right)}\times\nn
			&\big(28475597577033317899 D^3-90607604692549882710 D^2+\nn
			&107185135379627366777 D-85969113572839070718\big)\nn
			&+\frac{ s_{03} s_{01}^2}{72569239152670296 (D-3) \left(D^2-1\right)}\times\big(-5435043687063540225 D^3+\nn
			&34743610215547035316 D^2-133703649094040295147 D+\nn
			&138355148225398767416\big)+\frac{ s_{04} s_{01}^2}{473834443879200168 (D-3) \left(D^2-1\right)}\times\nn
			&\big(2027743034533980625 D^3-47755513091825901372 D^2+\nn
			&143945054906768879195 D-118509140188920876912\big)\nn
			&+\frac{ s_{02}^2 s_{01}}{4412994272797518 (D-3) \left(D^2-1\right)}\times\big(95273008629079876 D^3-\nn
			&5459007779794224849 D^2+34251474265548951092 D-68798575805871096567\big)\nn
			&+\frac{ s_{03}^2 s_{01}}{217925643101112 (D-3) \left(D^2-1\right)}\times\big(14312658358171485 D^3-\nn
			&308183885271005346 D^2+2120141865244853789 D-4345132269866178904\big)\nn
			&+\frac{ s_{04}^2 s_{01}}{9290871448611768 (D-3) \left(D^2-1\right)}\times\big(-8402381828589725 D^3-\nn
			&228340458232788810 D^2+2175040781006931191 D-4392791821329079248\big)\nn
			&+\frac{(20997586789 D-30664084565) s_{00} s_{01}}{1906477614 (D-3) (D-1)}+\frac{ s_{02} s_{03} s_{01}}{980665393955004 (D-3) \left(D^2-1\right)}\times\nn
			&\big(-178531580572981155 D^3+2760744792435584546 D^2-\nn
			&16899392445525068877 D+34463496611388041614\big)\nn
			&+\frac{ s_{02} s_{04} s_{01}}{6403168160529732 (D-3) \left(D^2-1\right)}\times\big(44888698974740075 D^3-\nn
			&2281481676512594142 D^2+17131774468846926949 D-34844020585587415026\big)\nn
			&+\frac{ s_{03} s_{04} s_{01}}{355731564473874 (D-3) \left(D^2-1\right)}\times\big(134329873732898840 D^2-\nn
			&1075850083783373373 D+2183705381853832549\big)\nn
			&+\frac{ s_{02}^3}{178905173221521 (D-3) \left(D^2-1\right)}\times\big(10367549408524951 D^3-\nn
			&244395249217448376 D^2+2493217982690141465 D-6695874076061568024\big)\nn
			&-\frac{ s_{03}^3}{654431360664 (D-3) \left(D^2-1\right)}\times\big(12563664070467 D^3-480258600687861 D^2+\nn
			&11117705113357806 D-37576216670725964\big)+\frac{(286248073-76106817 D) s_{00} s_{03}}{5725158 (D-3) (D-1)}\nn
			&+\frac{ s_{04}^3}{9290871448611768 (D-3) \left(D^2-1\right)}\times\big(591889764258953 D^3+\nn
			&18615866319928209 D^2+438017795639564794 D-2088530127407051460\big)\nn
			&+\frac{ s_{02} s_{03}^2}{5889882245976 (D-3) \left(D^2-1\right)}\times\big(470145534426423 D^3-\nn
			&16661705329346940 D^2+279904962791498257 D-883445043516938732\big)\nn
			&+\frac{ s_{02} s_{04}^2}{251104633746264 (D-3) \left(D^2-1\right)}\times\big(-186005811461743 D^3-\nn
			&122956210936788 D^2+238775662666263895 D-941769244106471412\big)\nn
			&+\frac{\left(-378033081851837 D^2-28380809735275779 D+119557455493102352\right) s_{03} s_{04}^2}{27900514860696 (D-3) \left(D^2-1\right)}\nn
			&+\frac{ s_{02}^2 s_{03}}{53008940213784 (D-3) \left(D^2-1\right)}\times\big(-5864446929424329 D^3+\nn
			&183078746487862324 D^2-2359327728622119111 D+6895583935338277052\big)\nn
			&+\frac{s_{02}^2 s_{04}}{346117197866472 (D-3) \left(D^2-1\right)}\times\big(993713336121961 D^3-\nn
			&104743917618759156 D^2+2191839897969601319 D-7146277328453200380\big)\nn
			& +\frac{\left(-467242505501283 D^2+30910888602825155 D-116182101038199856\right) s_{03}^2 s_{04}}{4273051825512 (D-3) \left(D^2-1\right)}\nn
			&+\frac{\left(1961957761174202 D^2-64920756785680566 D+228237749746197820\right) s_{02} s_{03} s_{04}}{4807183303701 (D-3) \left(D^2-1\right)}\nn
			&+\frac{(64066921 D-274208177) s_{00} s_{04}}{37381914 (D-3) (D-1)}+\frac{(268004836 D-1108569860) s_{00} s_{02}}{25763211 (D-3) (D-1)}\co
		\end{align}
		\begin{align}
			F^{(3)}_{5;\what{234}}=&\frac{ s_{01}^3}{16806910847891893632 \left(D^2-1\right)}\times\big(45979283969009274300559 D^2+\nn
			&1047032357312884852901 D-46656230678599317469812\big)\nn
			&+\frac{ s_{02} s_{01}^2}{33767531499168 \left(D^2-1\right)}\times
			\big(-74104146458248832 D^2+\nn
			&29506403105945531 D+36745597196241525\big)\nn
			&+\frac{ s_{03} s_{01}^2}{1770919429734144 \left(D^2-1\right)}\times\big(-900085462820883747 D^2-\nn
			&1780257440638046161 D+3064860211892055028\big)\nn
			&+\frac{\left(16379947016073091 D^2+598222244103749657 D-688591797903427356\right) s_{04} s_{01}^2}{3723697983359232 \left(D^2-1\right)}\nn
			&+\frac{\left(-46832281406738 D^2+487441428989732 D-919312210645983\right) s_{02}^2 s_{01}}{114079498308 \left(D^2-1\right)}\nn
			&+\frac{\left(-5962412169596571 D^2+57241427343714135 D-108319299169698748\right) s_{03}^2 s_{01}}{10636152731136 \left(D^2-1\right)}\nn
			&+\frac{\left(-9882296771733875 D^2+317362498527208511 D-611771709694482252\right) s_{04}^2 s_{01}}{2482465322239488 \left(D^2-1\right)}\nn
			&+\frac{\left(188034785817840 D^2-970303447937245 D+1748833608329821\right) s_{02} s_{03} s_{01}}{101403998496 \left(D^2-1\right)}\nn
			&+\frac{\left(-114853210406800 D^2+975040680624469 D-1817277951053637\right) s_{02} s_{04} s_{01}}{662108460768 \left(D^2-1\right)}\nn
			&+\frac{\left(66634569689337 D^2-976812872646037 D+1863724399749532\right) s_{03} s_{04} s_{01}}{588540854016 \left(D^2-1\right)}\nn
			&+\frac{\left(-2271838998767 D^2+36591900808730 D-87535035935325\right) s_{02}^3}{4624844526 \left(D^2-1\right)}\nn
			&+\frac{\left(11671435834413 D^2-1899510805030945 D+5546345527961140\right) s_{03}^3}{191642391552 \left(D^2-1\right)}\nn
			&+\frac{\left(16098192433184611 D^2+144789759004827209 D-565364111032193004\right) s_{04}^3}{4964930644478976 \left(D^2-1\right)}\nn
			&+\frac{\left(-990342600288 D^2+33466366766187 D-91942066376347\right) s_{02} s_{03}^2}{1218066048 \left(D^2-1\right)}\nn
			&+\frac{\left(16181230755488 D^2+500766696393835 D-1664193721989627\right) s_{02} s_{04}^2}{882811281024 \left(D^2-1\right)}\nn
			&+\frac{\left(-31681674345033 D^2-481937156164675 D+1698523705808332\right) s_{03} s_{04}^2}{784721138688 \left(D^2-1\right)}\nn
			&+\frac{\left(772076303514 D^2-17465983786156 D+44961327781741\right) s_{02}^2 s_{03}}{685162152 \left(D^2-1\right)}\nn
			&+\frac{\left(-317817093958 D^2+16432322501716 D-46449248275737\right) s_{02}^2 s_{04}}{4473705816 \left(D^2-1\right)}\nn
			&+\frac{\left(829966618209 D^2+30291424183827 D-96937318177204\right) s_{03}^2 s_{04}}{7069559808 \left(D^2-1\right)}\nn
			&+\frac{(94917907364339-31480868378483 D) s_{02} s_{03} s_{04}}{3976627392 \left(D^2-1\right)}-\frac{115943 s_{00} s_{01}}{49284 (D-1)}\nn
			&-\frac{5041 s_{00} s_{02}}{333 (D-1)}+\frac{5041 s_{00} s_{03}}{296 (D-1)}-\frac{85697 s_{00} s_{04}}{32856 (D-1)}\co
		\end{align}
		\bea
		F^{(3)}_{5;\what{1234}}=\frac{25411681 (D-2) \left(46 s_{01}+296 s_{02}-333 s_{03}+51 s_{04}\right){}^3}{13384745856 (D+1)}\ed
		\eea
	\end{subequations}
	These results have been checked with FIRE6 \cite{Smirnov:2019qkx}.

	\section{Divergence of reduction coefficients of two-loop sunset topology }
	
	In this part, we give the related $C_i,C_j$ coefficients for rank level 3 with the chosen
	tensor structure, which have been used to solve ${\cal B}$ by \eref{finite-1}. 
	
	\label{appdix:sunset}
    \begin{itemize}
    	\item The pole of $s_{11}^{-3}$
    	\allowdisplaybreaks
	\begin{align}
		\left( \begin{array}{c} C^{(3,0)}_2 \\ C^{(3,0)}_3 \\ C^{(3,0)}_4 \end{array}\right)\Bigg\vert_{s_{11}^{-3}}&=\left(
		\begin{array}{c}
			\frac{(D+2) \left(4 (1-2 D) M_2^2 M_{12,3}+(3 D-2) M_{12,3}^2+8 (D-1) M_1^2 M_2^2\right) s_{01}^3}{(D-1) (3 D-2)} \\
			\frac{2 (D+2) M_1^2 \left((5 D-4) M_{12,3}+4 (1-2 D) M_2^2\right) s_{01}^3}{(D-1) (3 D-2)} \\
			-\frac{2 (D+2) \left((4 D-2) M_{12,3}+(D-2) M_1^2+4 (1-2 D) M_2^2\right) s_{01}^3}{(D-1) (3 D-2)} \\
		\end{array}
		\right)\co
		\nn
		\left( \begin{array}{c} C^{(2,1)}_2 \\ C^{(2,1)}_3 \\ C^{(2,1)}_4 \end{array}\right)\Bigg\vert_{s_{11}^{-3}}&=\left(
		\begin{array}{c}
			\frac{2 (D+2) M_2^2 \left((2 D-1) M_{12,3}-2 (D-1) M_1^2\right) s_{01}^2 s_{0'1}}{(D-1) (3 D-2)} \\
			-\frac{2 (D+2) M_1^2 \left((D-1) M_{12,3}+2 (1-2 D) M_2^2\right) s_{01}^2 s_{0'1}}{(D-1) (3 D-2)} \\
			\frac{(D+2) \left(D M_{12,3}+4 (D-1) M_1^2+4 (1-2 D) M_2^2\right) s_{01}^2 s_{0'1}}{(D-1) (3 D-2)} \\
		\end{array}
		\right)\co
		\nn
		\left( \begin{array}{c} C^{(1,2)}_2 \\ C^{(1,2)}_3 \\ C^{(1,2)}_4 \end{array}\right)\Bigg\vert_{s_{11}^{-3}}&=\left(
		\begin{array}{c}
			\frac{2 (D+2) M_2^2 \left((4 D-2) M_1^2-(D-1) M_{12,3}\right) s_{01} s_{0'1}^2}{(D-1) (3 D-2)} \\
			-\frac{2 (D+2) M_1^2 \left((1-2 D) M_{12,3}+2 (D-1) M_2^2\right) s_{01} s_{0'1}^2}{(D-1) (3 D-2)} \\
			-\frac{(D+2) \left(-D M_{12,3}+(8 D-4) M_1^2-4 (D-1) M_2^2\right) s_{01} s_{0'1}^2}{(D-1) (3 D-2)} \\
		\end{array}
		\right)\co
		\nn
		\left( \begin{array}{c} C^{(0,3)}_2 \\ C^{(0,3)}_3 \\ C^{(0,3)}_4 \end{array}\right)\Bigg\vert_{s_{11}^{-3}}&=\left(
		\begin{array}{c}
			-\frac{2 (D+2) M_2^2 \left((4-5 D) M_{12,3}+(8 D-4) M_1^2\right) s_{0'1}^3}{(D-1) (3 D-2)} \\
			\frac{(D+2) \left(4 M_1^2 \left((1-2 D) M_{12,3}+2 (D-1) M_2^2\right)+(3 D-2) M_{12,3}^2\right) s_{0'1}^3}{(D-1) (3 D-2)} \\
			\frac{2 (D+2) \left(2 (1-2 D) M_{12,3}+(8 D-4) M_1^2-(D-2) M_2^2\right) s_{0'1}^3}{(D-1) (3 D-2)} \\
		\end{array}
		\right)\ed
	\end{align}
	\begin{align}
			\left( \begin{array}{c} C^{(r_1,r_2)}_1 \\ C^{(r_1,r_2)}_5 \\ C^{(r_1,r_2)}_6\\
			C^{(r_1,r_2)}_7 \end{array}\right)\Bigg\vert_{s_{11}^{-3}}&=\left(
		\begin{array}{c}
		0 \\
		0 \\
		0 \\
		0 \\
	\end{array}
   \right),\forall r_1+r_2=3\ed
	\end{align}
	\item The pole of $s_{11}^{-2}$
	\begin{align}
		\left( \begin{array}{c} C^{(3,0)}_2 \\ C^{(3,0)}_3 \\ C^{(3,0)}_4 \end{array}\right)\Bigg\vert_{s_{01}^3s_{11}^{-2}}&=\left(
		\begin{array}{c}
			\frac{4 \left(\left(-3 D^2-5 D+2\right) M_3^2+D (5 D-2) M_1^2-D (D+2) M_2^2\right)}{(D-1) (3 D-2)} \\
			\frac{2 (D+2) \left(2 (5 D-4) M_1^2+(4-5 D) M_2^2-3 D M_3^2\right)}{(D-1) (3 D-2)} \\
			-\frac{2 \left(5 D^2+6 D-8\right)}{3 D^2-5 D+2} \\
		\end{array}
		\right)\co
		\nn
		\left( \begin{array}{c} C^{(2,1)}_2 \\ C^{(2,1)}_3 \\ C^{(2,1)}_4 \end{array}\right)\Bigg\vert_{s_{01}^2s_{0'1}s_{11}^{-2}}&=\left(
		\begin{array}{c}
			-\frac{2 (D+2) \left((2 D-1) M_1^2+(2-4 D) M_2^2-M_3^2\right)}{(D-1) (3 D-2)} \\
			\frac{2 \left(\left(D^2-16 D+12\right) M_1^2+(D+2) \left((4 D-3) M_2^2+M_3^2\right)\right)}{(D-1) (3 D-2)} \\
			-\frac{D^2-10 D+8}{3 D^2-5 D+2} \\
		\end{array}
		\right)\co
		\nn
		\left( \begin{array}{c} C^{(1,2)}_2 \\ C^{(1,2)}_3 \\ C^{(1,2)}_4 \end{array}\right)\Bigg\vert_{s_{01}s_{0'1}^2s_{11}^{-2}}&=\left(
		\begin{array}{c}
			\frac{2 \left(\left(4 D^2+5 D-6\right) M_1^2+\left(D^2-16 D+12\right) M_2^2+(D+2) M_3^2\right)}{(D-1) (3 D-2)} \\
			\frac{2 (D+2) \left((4 D-2) M_1^2+(1-2 D) M_2^2+M_3^2\right)}{(D-1) (3 D-2)} \\
			-\frac{D^2-10 D+8}{3 D^2-5 D+2} \\
		\end{array}
		\right)\co
		\nn
		\left( \begin{array}{c} C^{(0,3)}_2 \\ C^{(0,3)}_3 \\ C^{(0,3)}_4 \end{array}\right)\Bigg\vert_{s_{0'1}^3s_{11}^{-2}}&=\left(
		\begin{array}{c}
			-\frac{2 (D+2) \left((5 D-4) M_1^2+(8-10 D) M_2^2+3 D M_3^2\right)}{(D-1) (3 D-2)} \\
			-\frac{4 \left(\left(3 D^2+5 D-2\right) M_3^2+D (D+2) M_1^2+(2-5 D) D M_2^2\right)}{(D-1) (3 D-2)} \\
			-\frac{2 \left(5 D^2+6 D-8\right)}{3 D^2-5 D+2} \\
		\end{array}
		\right)\ed
	\end{align}
    \begin{align}
        \left( \begin{array}{c} C^{(3,0)}_1 \\ C^{(3,0)}_5 \\ C^{(3,0)}_6\\C^{(3,0)}_7 \end{array}\right)\Bigg\vert_{s_{01}^3s_{11}^{-2}}&=\left(
        \begin{array}{c}
        	-\frac{2 (D+2) \left((7 D-4) M_1^4+2 M_1^2 \left((D-4) M_2^2+(2-5 D) M_3^2\right)-\left(M_2^2-M_3^2\right) \left((5 D-4) M_2^2+3 D M_3^2\right)\right)}{(D-1) (3 D-2)} \\
        	-\frac{4 (D+2) (2 D-1) \left(D M_1^2-2 \left(M_2^2+M_3^2\right)\right)}{(D-1) D (3 D-2)} \\
        	-\frac{2 (D+2) \left(\left(3 D^2+2 D-4\right) M_1^2+D \left((4-5 D) M_2^2-3 D M_3^2\right)\right)}{(D-1) D (3 D-2)} \\
        	\frac{2 (D+2) \left(\left(7 D^2-10 D+4\right) M_1^2+D \left((4-5 D) M_2^2-3 D M_3^2\right)\right)}{(D-1) D (3 D-2)} \\
        \end{array}
        \right)\co\nn
        \left( \begin{array}{c} C^{(2,1)}_1 \\ C^{(2,1)}_5 \\ C^{(2,1)}_6\\C^{(2,1)}_7 \end{array}\right)\Bigg\vert_{s_{01}^2s_{0'1}s_{11}^{-2}}&=\left(
        \begin{array}{c}
        	\frac{2 (D+2) \left((2 D-1) M_1^4-2 M_1^2 \left((D+1) M_2^2+D M_3^2\right)-\left(M_2^2-M_3^2\right) \left((4 D-3) M_2^2+M_3^2\right)\right)}{(D-1) (3 D-2)} \\
        	\frac{2 (D+2) \left(D (2 D-1) M_1^2+(4-7 D) M_2^2-D M_3^2\right)}{(D-1) D (3 D-2)} \\
        	-\frac{2 (D+2) \left((4-5 D) M_1^2+D \left((4 D-3) M_2^2+M_3^2\right)\right)}{(D-1) D (3 D-2)} \\
        	-\frac{2 (D+2) \left((2 D-1) M_1^2+(3-4 D) M_2^2-M_3^2\right)}{(D-1) (3 D-2)} \\
        \end{array}
        \right)\co\nn
        \left( \begin{array}{c} C^{(1,2)}_1 \\ C^{(1,2)}_5 \\ C^{(1,2)}_6\\C^{(1,2)}_7 \end{array}\right)\Bigg\vert_{s_{01}s_{0'1}^2s_{11}^{-2}}&=\left(
        \begin{array}{c}
        	-\frac{2 (D+2) \left((4 D-3) M_1^4+2 M_1^2 \left((D+1) M_2^2-2 (D-1) M_3^2\right)+(1-2 D) M_2^4+2 D M_2^2 M_3^2-M_3^4\right)}{(D-1) (3 D-2)} \\
        	-\frac{2 (D+2) \left(D (4 D-3) M_1^2+(4-5 D) M_2^2+D M_3^2\right)}{(D-1) D (3 D-2)} \\
        	-\frac{2 (D+2) \left((7 D-4) M_1^2+D \left((1-2 D) M_2^2+M_3^2\right)\right)}{(D-1) D (3 D-2)} \\
        	\frac{2 (D+2) \left((4 D-3) M_1^2+(1-2 D) M_2^2+M_3^2\right)}{(D-1) (3 D-2)} \\
        \end{array}
        \right)\co\nn
        \left( \begin{array}{c} C^{(0,3)}_1 \\ C^{(0,3)}_5 \\ C^{(0,3)}_6\\C^{(0,3)}_7 \end{array}\right)\Bigg\vert_{s_{0'1}^3s_{11}^{-2}}&=\left(
        \begin{array}{c}
        	\frac{2 (D+2) \left((5 D-4) M_1^4-2 M_1^2 \left((D-4) M_2^2+(D-2) M_3^2\right)-\left(M_2^2-M_3^2\right) \left((7 D-4) M_2^2-3 D M_3^2\right)\right)}{(D-1) (3 D-2)} \\
        	\frac{2 (D+2) \left(\left(-3 D^2-2 D+4\right) M_2^2+3 D^2 M_3^2+D (5 D-4) M_1^2\right)}{(D-1) D (3 D-2)} \\
        	-\frac{4 (D+2) (2 D-1) \left(D M_2^2-2 M_1^2-2 M_3^2\right)}{(D-1) D (3 D-2)} \\
        	-\frac{2 (D+2) \left(\left(-7 D^2+10 D-4\right) M_2^2+3 D^2 M_3^2+D (5 D-4) M_1^2\right)}{(D-1) D (3 D-2)} \\
        \end{array}
        \right)\ed
    \end{align}
	\item The pole of $s_{11}^{-1}$
	\begin{align}
		\left( \begin{array}{c} C^{(3,0)}_2 \\ C^{(3,0)}_3 \\ C^{(3,0)}_4 \end{array}\right)\Bigg\vert_{s_{01}^3s_{11}^{-1}}&=\left(
		\begin{array}{c}
			\frac{13 D^2+16 D-20}{(D-1) (3 D-2)} \\
			\frac{2 \left(5 D^2+6 D-8\right)}{(D-1) (3 D-2)} \\
			0 \\
		\end{array}
		\right)\co
		\nn
		\left( \begin{array}{c} C^{(2,1)}_2 \\ C^{(2,1)}_3 \\ C^{(2,1)}_4 \end{array}\right)\Bigg\vert_{s_{01}^2s_{0'1}s_{11}^{-1}}&=\left(
		\begin{array}{c}
			\frac{2 \left(2 D^2-9 D+6\right)}{3 D^2-5 D+2} \\
			\frac{2 \left(2 D^2-9 D+6\right)}{3 D^2-5 D+2} \\
			0 \\
		\end{array}
		\right)\co
		\nn
		\left( \begin{array}{c} C^{(1,2)}_2 \\ C^{(1,2)}_3 \\ C^{(1,2)}_4 \end{array}\right)\Bigg\vert_{s_{01}s_{0'1}^2s_{11}^{-1}}&=\left(
		\begin{array}{c}
			\frac{2 \left(2 D^2-9 D+6\right)}{3 D^2-5 D+2} \\
			\frac{2 \left(2 D^2-9 D+6\right)}{3 D^2-5 D+2} \\
			0 \\
		\end{array}
		\right)\co
		\nn
		\left( \begin{array}{c} C^{(0,3)}_2 \\ C^{(0,3)}_3 \\ C^{(0,3)}_4 \end{array}\right)\Bigg\vert_{s_{0'1}^3s_{11}^{-1}}&=\left(
		\begin{array}{c}
			\frac{2 \left(5 D^2+6 D-8\right)}{(D-1) (3 D-2)} \\
			\frac{13 D^2+16 D-20}{(D-1) (3 D-2)} \\
			0 \\
		\end{array}
		\right)\ed
	\end{align}
	\begin{align}
		\left( \begin{array}{c} C^{(3,0)}_1 \\ C^{(3,0)}_5 \\ C^{(3,0)}_6\\C^{(3,0)}_7 \end{array}\right)\Bigg\vert_{s_{01}^3s_{11}^{-1}}&=\left(
		\begin{array}{c}
			-\frac{8 (D+2) \left((3 D-2) M_1^2+(1-2 D) M_3^2\right)}{(D-1) (3 D-2)} \\
			\frac{8 D^2-60 D+40}{3 D^2-5 D+2} \\
			-\frac{2 \left(5 D^2+6 D-8\right)}{3 D^2-5 D+2} \\
			\frac{2 \left(5 D^2+6 D-8\right)}{(D-1) (3 D-2)} \\
		\end{array}
		\right)\co\nn
		\left( \begin{array}{c} C^{(2,1)}_1 \\ C^{(2,1)}_5 \\ C^{(2,1)}_6\\C^{(2,1)}_7 \end{array}\right)\Bigg\vert_{s_{01}^2s_{0'1}s_{11}^{-1}}&=\left(
		\begin{array}{c}
			\frac{2 \left(2 \left(D^2-5 D+2\right) M_3^2+4 (3 D-2) M_1^2+2 (2-3 D) D M_2^2\right)}{(D-1) (3 D-2)} \\
			-\frac{2 \left(2 D^2-9 D+6\right)}{3 D^2-5 D+2} \\
			-\frac{2 \left(2 D^2-9 D+6\right)}{3 D^2-5 D+2} \\
			\frac{2 \left(2 D^2-9 D+6\right)}{3 D^2-5 D+2} \\
		\end{array}
		\right)\co\nn
		\left( \begin{array}{c} C^{(1,2)}_1 \\ C^{(1,2)}_5 \\ C^{(1,2)}_6\\C^{(1,2)}_7 \end{array}\right)\Bigg\vert_{s_{01}s_{0'1}^2s_{11}^{-1}}&=\left(
		\begin{array}{c}
			\frac{2 \left(2 \left(D^2-5 D+2\right) M_3^2+2 (2-3 D) D M_1^2+4 (3 D-2) M_2^2\right)}{(D-1) (3 D-2)} \\
			-\frac{2 \left(2 D^2-9 D+6\right)}{3 D^2-5 D+2} \\
			-\frac{2 \left(2 D^2-9 D+6\right)}{3 D^2-5 D+2} \\
			\frac{2 \left(2 D^2-9 D+6\right)}{3 D^2-5 D+2} \\
		\end{array}
		\right)\co\nn
		\left( \begin{array}{c} C^{(0,3)}_1 \\ C^{(0,3)}_5 \\ C^{(0,3)}_6\\C^{(0,3)}_7 \end{array}\right)\Bigg\vert_{s_{0'1}^3s_{11}^{-1}}&=\left(
		\begin{array}{c}
			-\frac{8 (D+2) \left((3 D-2) M_2^2+(1-2 D) M_3^2\right)}{(D-1) (3 D-2)} \\
			-\frac{2 \left(5 D^2+6 D-8\right)}{3 D^2-5 D+2} \\
			\frac{8 D^2-60 D+40}{3 D^2-5 D+2} \\
			\frac{2 \left(5 D^2+6 D-8\right)}{(D-1) (3 D-2)} \\
		\end{array}
		\right)\ed
	\end{align}
\end{itemize}
	\appendix

	\bibliographystyle{JHEP}
	\bibliography{reference}

\providecommand{\href}[2]{#2}\begingroup\raggedright\begin{thebibliography}{10}

\bibitem{Feng:2021enk}
B.~Feng, T.~Li, and X.~Li, {\it {Analytic tadpole coefficients of one-loop
  integrals}},  {\em JHEP} {\bf 09} (2021) 081,
  [\href{http://arxiv.org/abs/2107.03744}{{\tt arXiv:2107.03744}}].

\bibitem{Hu:2021nia}
C.~Hu, T.~Li, and X.~Li, {\it {One-loop Feynman Integral Reduction by
  Differential Operators}},  \href{http://arxiv.org/abs/2108.00772}{{\tt
  arXiv:2108.00772}}.

\bibitem{Feng:2022uqp}
B.~Feng, T.~Li, H.~Wang, and Y.~Zhang, {\it {Reduction of General One-loop
  Integrals Using Auxiliary Vector}},
  \href{http://arxiv.org/abs/2203.14449}{{\tt arXiv:2203.14449}}.

\bibitem{Brown:1952eu}
L.~M. Brown and R.~P. Feynman, {\it {Radiative corrections to Compton
  scattering}},  {\em Phys. Rev.} {\bf 85} (1952) 231--244.

\bibitem{Melrose:1965kb}
D.~B. Melrose, {\it {Reduction of Feynman diagrams}},  {\em Nuovo Cim.} {\bf
  40} (1965) 181--213.

\bibitem{Passarino:1978jh}
G.~Passarino and M.~J.~G. Veltman, {\it {One Loop Corrections for e+ e-
  Annihilation Into mu+ mu- in the Weinberg Model}},  {\em Nucl. Phys. B} {\bf
  160} (1979) 151--207.

\bibitem{tHooft:1978jhc}
G.~'t~Hooft and M.~J.~G. Veltman, {\it {Scalar One Loop Integrals}},  {\em
  Nucl. Phys. B} {\bf 153} (1979) 365--401.

\bibitem{vanNeerven:1983vr}
W.~L. van Neerven and J.~A.~M. Vermaseren, {\it {LARGE LOOP INTEGRALS}},  {\em
  Phys. Lett. B} {\bf 137} (1984) 241--244.

\bibitem{Stuart:1987tt}
R.~G. Stuart, {\it {Algebraic Reduction of One Loop Feynman Diagrams to Scalar
  Integrals}},  {\em Comput. Phys. Commun.} {\bf 48} (1988) 367--389.

\bibitem{vanOldenborgh:1989wn}
G.~J. van Oldenborgh and J.~A.~M. Vermaseren, {\it {New Algorithms for One Loop
  Integrals}},  {\em Z. Phys. C} {\bf 46} (1990) 425--438.

\bibitem{Bern:1992em}
Z.~Bern, L.~J. Dixon, and D.~A. Kosower, {\it {Dimensionally regulated one loop
  integrals}},  {\em Phys. Lett. B} {\bf 302} (1993) 299--308,
  [\href{http://arxiv.org/abs/hep-ph/9212308}{{\tt hep-ph/9212308}}]. [Erratum:
  Phys.Lett.B 318, 649 (1993)].

\bibitem{Bern:1993kr}
Z.~Bern, L.~J. Dixon, and D.~A. Kosower, {\it {Dimensionally regulated pentagon
  integrals}},  {\em Nucl. Phys. B} {\bf 412} (1994) 751--816,
  [\href{http://arxiv.org/abs/hep-ph/9306240}{{\tt hep-ph/9306240}}].

\bibitem{Fleischer:1999hq}
J.~Fleischer, F.~Jegerlehner, and O.~V. Tarasov, {\it {Algebraic reduction of
  one loop Feynman graph amplitudes}},  {\em Nucl. Phys. B} {\bf 566} (2000)
  423--440, [\href{http://arxiv.org/abs/hep-ph/9907327}{{\tt hep-ph/9907327}}].

\bibitem{Binoth:1999sp}
T.~Binoth, J.~P. Guillet, and G.~Heinrich, {\it {Reduction formalism for
  dimensionally regulated one loop N point integrals}},  {\em Nucl. Phys. B}
  {\bf 572} (2000) 361--386, [\href{http://arxiv.org/abs/hep-ph/9911342}{{\tt
  hep-ph/9911342}}].

\bibitem{Denner:2002ii}
A.~Denner and S.~Dittmaier, {\it {Reduction of one loop tensor five point
  integrals}},  {\em Nucl. Phys. B} {\bf 658} (2003) 175--202,
  [\href{http://arxiv.org/abs/hep-ph/0212259}{{\tt hep-ph/0212259}}].

\bibitem{Duplancic:2003tv}
G.~Duplancic and B.~Nizic, {\it {Reduction method for dimensionally regulated
  one loop N point Feynman integrals}},  {\em Eur. Phys. J. C} {\bf 35} (2004)
  105--118, [\href{http://arxiv.org/abs/hep-ph/0303184}{{\tt hep-ph/0303184}}].

\bibitem{Denner:2005nn}
A.~Denner and S.~Dittmaier, {\it {Reduction schemes for one-loop tensor
  integrals}},  {\em Nucl. Phys. B} {\bf 734} (2006) 62--115,
  [\href{http://arxiv.org/abs/hep-ph/0509141}{{\tt hep-ph/0509141}}].

\bibitem{Ellis:2007qk}
R.~K. Ellis and G.~Zanderighi, {\it {Scalar one-loop integrals for QCD}},  {\em
  JHEP} {\bf 02} (2008) 002, [\href{http://arxiv.org/abs/0712.1851}{{\tt
  arXiv:0712.1851}}].

\bibitem{Ossola_2007}
G.~Ossola, C.~G. Papadopoulos, and R.~Pittau, {\it Reducing full one-loop
  amplitudes to scalar integrals at the integrand level},  {\em Nuclear Physics
  B} {\bf 763} (Feb, 2007) 147--169.

\bibitem{Bern:1994cg}
Z.~Bern, L.~J. Dixon, D.~C. Dunbar, and D.~A. Kosower, {\it {Fusing gauge
  theory tree amplitudes into loop amplitudes}},  {\em Nucl. Phys. B} {\bf 435}
  (1995) 59--101, [\href{http://arxiv.org/abs/hep-ph/9409265}{{\tt
  hep-ph/9409265}}].

\bibitem{Chetyrkin:1981qh}
K.~G. Chetyrkin and F.~V. Tkachov, {\it {Integration by Parts: The Algorithm to
  Calculate beta Functions in 4 Loops}},  {\em Nucl. Phys. B} {\bf 192} (1981)
  159--204.

\bibitem{Tkachov:1981wb}
F.~V. Tkachov, {\it {A Theorem on Analytical Calculability of Four Loop
  Renormalization Group Functions}},  {\em Phys. Lett. B} {\bf 100} (1981)
  65--68.

\bibitem{Ossola:2006us}
G.~Ossola, C.~G. Papadopoulos, and R.~Pittau, {\it {Reducing full one-loop
  amplitudes to scalar integrals at the integrand level}},  {\em Nucl. Phys. B}
  {\bf 763} (2007) 147--169, [\href{http://arxiv.org/abs/hep-ph/0609007}{{\tt
  hep-ph/0609007}}].

\bibitem{Ossola:2007bb}
G.~Ossola, C.~G. Papadopoulos, and R.~Pittau, {\it {Numerical evaluation of
  six-photon amplitudes}},  {\em JHEP} {\bf 07} (2007) 085,
  [\href{http://arxiv.org/abs/0704.1271}{{\tt arXiv:0704.1271}}].

\bibitem{Ellis:2007br}
R.~K. Ellis, W.~T. Giele, and Z.~Kunszt, {\it {A Numerical Unitarity Formalism
  for Evaluating One-Loop Amplitudes}},  {\em JHEP} {\bf 03} (2008) 003,
  [\href{http://arxiv.org/abs/0708.2398}{{\tt arXiv:0708.2398}}].

\bibitem{Bern:1994zx}
Z.~Bern, L.~J. Dixon, D.~C. Dunbar, and D.~A. Kosower, {\it {One loop n point
  gauge theory amplitudes, unitarity and collinear limits}},  {\em Nucl. Phys.
  B} {\bf 425} (1994) 217--260,
  [\href{http://arxiv.org/abs/hep-ph/9403226}{{\tt hep-ph/9403226}}].

\bibitem{Britto:2004nc}
R.~Britto, F.~Cachazo, and B.~Feng, {\it {Generalized unitarity and one-loop
  amplitudes in N=4 super-Yang-Mills}},  {\em Nucl. Phys. B} {\bf 725} (2005)
  275--305, [\href{http://arxiv.org/abs/hep-th/0412103}{{\tt hep-th/0412103}}].

\bibitem{Britto:2005ha}
R.~Britto, E.~Buchbinder, F.~Cachazo, and B.~Feng, {\it {One-loop amplitudes of
  gluons in SQCD}},  {\em Phys. Rev. D} {\bf 72} (2005) 065012,
  [\href{http://arxiv.org/abs/hep-ph/0503132}{{\tt hep-ph/0503132}}].

\bibitem{Campbell:1996zw}
J.~M. Campbell, E.~W.~N. Glover, and D.~J. Miller, {\it {One loop tensor
  integrals in dimensional regularization}},  {\em Nucl. Phys. B} {\bf 498}
  (1997) 397--442, [\href{http://arxiv.org/abs/hep-ph/9612413}{{\tt
  hep-ph/9612413}}].

\bibitem{Bern:1997sc}
Z.~Bern, L.~J. Dixon, and D.~A. Kosower, {\it {One loop amplitudes for e+ e- to
  four partons}},  {\em Nucl. Phys. B} {\bf 513} (1998) 3--86,
  [\href{http://arxiv.org/abs/hep-ph/9708239}{{\tt hep-ph/9708239}}].

\bibitem{Anastasiou:2006gt}
C.~Anastasiou, R.~Britto, B.~Feng, Z.~Kunszt, and P.~Mastrolia, {\it {Unitarity
  cuts and Reduction to master integrals in d dimensions for one-loop
  amplitudes}},  {\em JHEP} {\bf 03} (2007) 111,
  [\href{http://arxiv.org/abs/hep-ph/0612277}{{\tt hep-ph/0612277}}].

\bibitem{Britto:2010um}
R.~Britto and E.~Mirabella, {\it {Single Cut Integration}},  {\em JHEP} {\bf
  01} (2011) 135, [\href{http://arxiv.org/abs/1011.2344}{{\tt
  arXiv:1011.2344}}].

\bibitem{Anastasiou:2006jv}
C.~Anastasiou, R.~Britto, B.~Feng, Z.~Kunszt, and P.~Mastrolia, {\it
  {D-dimensional unitarity cut method}},  {\em Phys. Lett. B} {\bf 645} (2007)
  213--216, [\href{http://arxiv.org/abs/hep-ph/0609191}{{\tt hep-ph/0609191}}].

\bibitem{Britto:2006fc}
R.~Britto and B.~Feng, {\it {Unitarity cuts with massive propagators and
  algebraic expressions for coefficients}},  {\em Phys. Rev. D} {\bf 75} (2007)
  105006, [\href{http://arxiv.org/abs/hep-ph/0612089}{{\tt hep-ph/0612089}}].

\bibitem{Britto:2007tt}
R.~Britto and B.~Feng, {\it {Integral coefficients for one-loop amplitudes}},
  {\em JHEP} {\bf 02} (2008) 095, [\href{http://arxiv.org/abs/0711.4284}{{\tt
  arXiv:0711.4284}}].

\bibitem{Britto:2008vq}
R.~Britto, B.~Feng, and P.~Mastrolia, {\it {Closed-Form Decomposition of
  One-Loop Massive Amplitudes}},  {\em Phys. Rev. D} {\bf 78} (2008) 025031,
  [\href{http://arxiv.org/abs/0803.1989}{{\tt arXiv:0803.1989}}].

\bibitem{Britto:2008sw}
R.~Britto, B.~Feng, and G.~Yang, {\it {Polynomial Structures in One-Loop
  Amplitudes}},  {\em JHEP} {\bf 09} (2008) 089,
  [\href{http://arxiv.org/abs/0803.3147}{{\tt arXiv:0803.3147}}].

\bibitem{Feng:2013sa}
B.~Feng and H.~Wang, {\it {Analytic structure of one-loop coefficients}},  {\em
  JHEP} {\bf 05} (2013) 104, [\href{http://arxiv.org/abs/1301.7510}{{\tt
  arXiv:1301.7510}}].

\bibitem{Feng:2022rwj}
B.~Feng, J.~Gong, and T.~Li, {\it {Universal Treatment of Reduction for
  One-Loop Integrals in Projective Space}},
  \href{http://arxiv.org/abs/2204.03190}{{\tt arXiv:2204.03190}}.

\bibitem{Denner:1991kt}
A.~Denner, {\it {Techniques for calculation of electroweak radiative
  corrections at the one loop level and results for W physics at LEP-200}},
  {\em Fortsch. Phys.} {\bf 41} (1993) 307--420,
  [\href{http://arxiv.org/abs/0709.1075}{{\tt arXiv:0709.1075}}].

\bibitem{Feng:2022iuc}
B.~Feng and T.~Li, {\it {PV-Reduction of Sunset Topology with Auxiliary
  Vector}},  \href{http://arxiv.org/abs/2203.16881}{{\tt arXiv:2203.16881}}.

\bibitem{Smirnov:2019qkx}
A.~V. Smirnov and F.~S. Chuharev, {\it {FIRE6: Feynman Integral REduction with
  Modular Arithmetic}},  {\em Comput. Phys. Commun.} {\bf 247} (2020) 106877,
  [\href{http://arxiv.org/abs/1901.07808}{{\tt arXiv:1901.07808}}].

\end{thebibliography}\endgroup
	
\end{document}